\documentclass{aa}

\usepackage{color}
\usepackage{multirow}

\bibpunct{(}{)}{;}{a}{}{,}
\usepackage{graphicx}
\usepackage[varg]{txfonts}
\usepackage{amsmath}
\begin{document}

\title{Magnetic torques on T Tauri stars: accreting vs. non-accreting systems}

   \author{G. Pantolmos\inst{1},
          C. Zanni\inst{2},
          \and
           J. Bouvier\inst{1}
          }

   \institute{Univ. Grenoble Alpes, CNRS, IPAG, 38000 Grenoble, France\\
     \email{george.pantolmos@univ-grenoble-alpes.fr}
     \and
     INAF – Osservatorio Astrofisico di Torino, Strada Osservatorio
     20, 10025 Pino Torinese, Italy
   }


 
  \abstract
   {Classical T Tauri stars (CTTs) magnetically interact with their
     surrounding disks, a process that is thought to regulate their
     rotational evolution.}
   {We compute torques acting onto the stellar surface of CTTs arising
     from different accreting (accretion funnels) and ejecting
     (stellar winds and magnetospheric ejections) flow
     components. Besides, we compare the magnetic braking due to
     stellar winds in two different systems: isolated  (i.e.,
     weak-line T Tauri and main-sequence) and accreting (i.e.,
     classical T Tauri) stars.}
   {We use 2.5D magnetohydrodynamic, time-dependent, axisymmetric
     simulations, computed with the PLUTO code. For both systems the
     stellar wind is thermally driven. In the star-disk-interaction (SDI)
     simulations the accretion disk is Keplerian, viscous, and
     resistive, modeled with an alpha prescription. Two
     series of simulations are presented, one for each system (i.e.,
     isolated and accreting stars).}
   {In classical T Tauri systems the presence
     of magnetospheric ejections confines the stellar-wind expansion,
     resulting in a hourglass-shaped geometry of the outflow and
     the formation of the accretion columns modifies the amount 
     of open magnetic flux exploited by the stellar wind. These
     effects have a strong impact on the stellar wind properties
     and we show that the stellar wind braking is more efficient in
     the star-disk-interacting systems than in the isolated ones.  
     We further derive torque scalings, over a wide range of magnetic 
     field strengths, for each flow component in a
     star-disk-interacting system (i.e., magnetospheric
     accretion/ejections, stellar winds), which directly applies a
     torque on the stellar surface.}
  {In all the performed SDI simulations the stellar wind extracts less
    than 2\% of the mass accretion rate and the disk is truncated up
    to 66\% of the corotation radius. All simulations show a net
    spin-up torque. We conclude that in order to achieve a stellar-spin
    equilibrium we need either more massive stellar winds or disks
    being truncated closer to the corotation radius, which increases
    the torque efficiency of the magnetospheric ejections.}


   \maketitle
%

\section{Introduction}

Classical T Tauri stars (CTTs) are young stellar objects (few Myr
old), with $M_{\ast} \lesssim 2M_{\sun}$, surrounded by accretion
disks \citep[see e.g., review by][]{Hartmann:2016aa}. These stars are 
magnetically active exhibiting multipolar fields, typically $\sim$kG
strong \citep[e.g.,][]{Johns-Krull:2007aa,Donati:2008ac,Donati:2019aa,Donati:2020aa,Gregory:2008aa,Johnstone:2014ab} 
and show signatures of mass accretion, with $\dot{M}_{acc}$ varying
from $10^{-8}$ to $10^{-10} \mathrm{M}_{\sun}\ \mathrm{yr}^{-1}$
\citep[e.g.,][]{Gullbring:1998aa,Hartmann:1998aa,Herczeg:2008aa,Ingleby:2014aa,Venuti:2014aa,Alcala:2017aa}.
The measured stellar magnetic fields are strong enough to disrupt the
disk and channel the accretion flow into funnels that impact the
stellar surface at near free-fall speed, forming hot accretion spots.

Despite being in a phase of stellar contraction and accretion,
CTTs are observed to be slow rotators, with rotation periods of
$\lesssim$ 8 days, which corresponds to about (or less than) 10\% of
their break-up limit \citep{Herbst:2007aa,Bouvier:2014aa}. In
addition, rotation-period distributions from young stellar clusters
indicate that CTTs rotate at a constant spin rate with
time \citep{Rebull:2004aa,Irwin:2009aa,Gallet:2013aa,Gallet:2019aa,Amard:2016aa}.
The latter features suggest the presence of angular-momentum-loss 
mechanisms acting on CTTs, which prevent their stellar surfaces to
spin up due to both accretion and contraction.

The general consensus is that the angular momentum evolution of
  CTTs is controlled by the interaction of the stellar magnetic field
  with the surrounding accretion disk and its environment. The
  star-disk magnetospheric interaction and the associated outflows
  have been extensively studied in the literature, using either
  semi-analytic models \citep[e.g.,][and references
therein]{Ghosh:1979ab,Collier-Cameron:1993aa,Lovelace:1995aa, 
  Armitage:1996aa,Agapitou:2000aa,Ferreira:2006aa,Matt:2005ab,Mohanty:2008aa,
  Sauty:2011aa} or numerical simulations \citep[e.g.,][and references
therein]{Hayashi:1996aa,
  Goodson:1997aa,Miller:1997aa,Kuker:2003aa,Zanni:2009aa,Zanni:2013aa,
  Cemeljic:2013aa,Kulkarni:2013aa,Romanova:2013aa,Cemeljic:2019aa}. Within this scenario,
several mechanisms have been proposed to explain the
rotation periods of CTTs. One of the first star-disk interaction
models applied to CTTs was based on the scenario originally proposed
by \citet{Ghosh:1979ab} for X-Ray pulsars. In this model, the stellar
field maintains a connection with the disk outside the corotation
radius, where a Keplerian disk rotates slower than the star. As a
consequence, the star can transfer 
angular momentum to the disk along the field lines connecting the two
objects, with the disk rotation controlling the stellar rotation
\citep[i.e. the disk-locking mechanism,
see e.g.,][]{Armitage:1996aa,Agapitou:2000aa,Matt:2005ab,Zanni:2009aa}.
However, this mechanism requires an extended stellar
magnetosphere that connects to the disk over a broad region beyond the
corotation radius, which seems unlikely \citep[see e.g.,][]{Matt:2005ab}. 
Different types of outflows, removing the excess stellar angular
momentum from the star-disk system instead of transferring it back to
the disk, have since been proposed as an alternative solution.  
Disk winds could effectively remove disk's angular momentum so that the
magnetospheric accretion does not affect the stellar angular momentum 
evolution \citep[e.g., X-winds,][]{Shu:1994aa,Cai:2008aa}.
Other authors proposed the idea of accretion-powered stellar winds 
\citep[e.g.,][]{Matt:2005aa,Matt:2007aa} in which magnetospheric
accretion acts as an additional energy source to drive the
stellar outflow, increasing the mass loss rate and consequently 
the spin-down efficiency of these winds.
 Different studies attempted to model the concurrent presence of
  stellar and disk winds (as a two-component outflow) with the main
  purpose of investigating the large-scale properties of protostellar
  jets \citep[see
  e.g.,][]{Bogovalov:2001aa,Fendt:2009aa,Matsakos:2009aa}.  
Another class of outflows exploits magnetospheric field lines that
still connect the star to the disk through a quasi-periodic and
unsteady process of inflation and reconnection of these magnetic
surfaces. The stellar magnetic surfaces exploited by these outflows 
can interact with a disk magnetic field whose magnetic moment can be
aligned \citep[e.g., ReX-winds,][]{Ferreira:2000aa} or anti-aligned
\citep[e.g., conical winds or magnetospheric ejections;
see][]{Romanova:2009ab,Zanni:2013aa} with respect to the
stellar one. Since these ejections take advantage
of magnetic field lines still connecting the star with the disk, they
can exchange mass, energy and 
angular momentum with both of them. In particular the spin-down
efficiency of this class of outflows is strongly increased when the
disk is truncated close or beyond the disk corotation radius, so that
they can efficiently tap the stellar rotational energy
\citep[propeller regime][]{Romanova:2005aa,Romanova:2009ab,Ustyugova:2006aa,Zanni:2013aa}. 
On the other hand, the different models of the propeller regime
\citep[e.g.,][]{Romanova:2005aa,Romanova:2009ab,Ustyugova:2006aa,Zanni:2013aa}
tend to predict a strong accretion variability that does not seem to
have any observational counterpart.

The aim of this study is to investigate the torques exerted onto a
CTTs by different flow components of a star-disk interacting
system. We will parametrize these external torques and provide
formulae, which have been proven useful for studies that attempt to
simulate the rotational evolution of late-type stars
\citep[e.g.,][]{Gallet:2013aa,Gallet:2015aa,Gallet:2019aa,Johnstone:2015ad,
  Matt:2015aa,Amard:2016aa,Amard:2019aa,Sadeghi-Ardestani:2017aa,See:2017aa,Garraffo:2018aa}. 
Our results will be based on magneto-hydrodynamic (MHD),
time-dependent, axisymmetric numerical simulations of an accretion
disk interacting with the magnetosphere of a rotating star. The
accretion disk is taken to be initially Keplerian, viscous, and resistive, modelled with
an alpha prescription \citep{Shakura:1973aa}. Furthermore,
we introduce a new equation of state with a temperature-dependent
polytropic index, which allows to simulate at the same time a
quasi-isothermal stellar wind and an adiabatic disk.
Our simulations will model the different flow components
that directly apply a torque on the stellar surface: accretion
funnel flows that increase the stellar angular momentum; 
magnetized stellar winds that provide a spin-down torque; intermittent
magnetospheric ejections \citep[hereafter MEs,][]{Zanni:2013aa}, consequence
of the differential rotation between the star and the inner disk, which
can either spin up or spin down the stellar surface. In particular we
will focus on regimes where the disk truncation radius does not exceed
the corotation radius, providing a steady accretion flow.   
In addition, we present stellar wind simulations from isolated stars
\citep[e.g.,][]{Washimi:1993aa,Keppens:1999aa,Matt:2012ab,Reville:2015ab,Pantolmos:2017aa,Finley:2018aa},
which are used to compare the stellar-wind torque efficiency in the
two different systems (diskless stars vs. star-disk-interacting
systems). 

In Sect. \ref{sec_num_set} we discuss the
numerical method, the initial and boundary conditions of the
simulations. In Sect. \ref{sec_mhdsim}, we present the global
phenomenology of our numerical solutions, focusing on representative
examples of the two systems (i.e., star-disk interaction and isolated
stellar winds) simulated here. In Sect. \ref{sec_scallaws}, we present the
the main results of this study. In particular, in Sect. \ref{sec_sw_torq},
we derive new stellar-wind torque scalings appropriate for the
classical T Tauri phase of stellar evolution, in Sect. \ref{sec_acc_torq}
and Sect. \ref{sec_me_torq}, we present torque prescriptions due to
accretion and magnetospheric ejections, respectively. In section
\ref{sec_disc} we compare our results with previous studies from the 
literature and finally, in section \ref{sec_concl} we summarize the
conclusions of this work. The details about the equation of state employed in
this numerical work are provided in Appendix \ref{app_eos} and in
Appendix \ref{app_sw_torq} we derive the formulation of the
stellar-wind torque discussed in Sect. \ref{sec_sw_torq}.

\section{Numerical setup}

\subsection{MHD equations and numerical method}
\label{sec_num_set}

The models presented in this work are numerical solutions of the
magneto-hydrodynamic (MHD) system of equations, including viscous and
resistive effects. In Gaussian units these equations are: 
\begin{equation}
\begin{aligned}
\label{eq_MHD}
\frac{\partial \rho}{\partial t} & + \nabla\cdot\left(\rho \vec {\upsilon}\right) = 0 \\
\frac{\partial \rho\vec{\upsilon}}{\partial t} & + \nabla \cdot \left[ 
\rho \vec{\upsilon}\vec{\upsilon} +
\left( P + \frac{\vec{B}\cdot\vec{B}}{8\pi} \right)\vec{I}-
\frac{\vec{B}\vec{B}}{4\pi} - \mathcal{T}
\right]  = \rho \vec{g} \\
\frac{\partial E}{\partial t} & + \nabla\cdot\left[
\left(E + P + \frac{\vec{B}\cdot\vec{B}}{8\pi}\right)\vec{\upsilon}-
\frac{\left(\vec{\upsilon}\cdot\vec{B}\right)\vec{B}}{4\pi} \right] =  \\
& = \rho \vec{g} \cdot \vec{\upsilon} + 
\left( \nabla \cdot \mathcal{T} \right) \cdot \vec{\upsilon}
- \frac{\vec{B}}{4\pi} \cdot \left(\nabla \times \eta_{\mathrm{m}} \vec{J}\right) \\
\frac{\partial \vec{B}}{\partial t} & + \nabla \times \left(\vec{B}\times\vec{\upsilon} + \eta_{\mathrm{m}} \vec{J} \right)= 0 \; . 
\end{aligned}
\end{equation}
The system of Eqs. (\ref{eq_MHD}) consists of mass, momentum and
energy conservation equations coupled to the induction equation to
follow the evolution of the magnetic field. We indicate with $\rho$
the mass density, $P$ the plasma thermal pressure, $\vec{B}$ and
$\vec{\upsilon}$ the magnetic field and velocity vectors respectively
and $\vec{I}$ the identity tensor. The total energy $E$ is the sum of
internal, kinetic and magnetic energy 
\begin{equation}
  E = \rho u +\rho \frac{\vec{\upsilon} \cdot \vec{\upsilon}}{2}
  +\frac{\vec{B} \cdot \vec{B}}{8 \pi} \; ,
  \label{eq_totenrg}
\end{equation}
where the definition of the specific internal energy $u(T)$ as a
function of temperature $T$ is provided in Appendix \ref{app_eos}. For
this work we have specifically developed a caloric equation of state
of a calorically imperfect gas, i.e. whose specific heats and their
ratio $\gamma$ are temperature-dependent. 
In particular the plasma in our models will behave almost isothermally
at high temperature and adiabatically at low temperature, so that we
will be able to simulate at the same time quasi-isothermal hot stellar
winds and cold adiabatic accretion disks. We set the equation of state
so that $\gamma = 1.05$ for $P/\rho > 0.1 \, GM_\ast/R_\ast$ and
$\gamma = 5/3$ for $P/\rho < 0.01 \, GM_\ast/R_\ast$, where $G$ is
Newton's gravitational constant, $M_\star$ and $R_\star$ are the
stellar mass and radius. 
The stellar gravitational acceleration is given by $\vec{g} =
-(GM_{\ast}/R^2) \hat{R}$. The electric current is defined by the
Amp\`ere's law, $\vec{J} = \nabla \times \vec{B}/4 \pi$. We indicate with
$\eta_{\mathrm{m}}$ and $\nu_{\mathrm{m}} = \eta_{\mathrm{m}}/4 \pi$
the magnetic resistivity and diffusivity respectively. 
The viscous stress tensor $\mathcal{T}$ is
\begin{equation}
  \mathcal{T} = \eta_{{\mathrm{v}}} \left[(\nabla \vec{\upsilon}) + (\nabla
  \vec{\upsilon})^T - \frac{2}{3} (\nabla \cdot \vec{\upsilon}) \vec{I}
  \right],
  \label{eq_Vstress}
\end{equation}
where $\eta_{{\mathrm{v}}}$ and $\nu_{\mathrm{v}} =
\eta_{{\mathrm{v}}}/\rho$ are the dynamic and the kinematic viscosity
respectively. 

Notice that the viscous and magnetic diffusive terms in the right hand
side of the total energy equation in the system of Eqs. (\ref{eq_MHD})
correspond to the work of the viscous forces and the diffusion of the
magnetic energy only. The dissipative viscous and Ohmic heating terms
are not included to avoid, in particular, a runaway irreversible
heating of the accretion disk, that would happen since no cooling
radiative effects have been taken into account.  
In the absence of viscosity and resistivity, the system of Eqs.
(\ref{eq_MHD}) reduces to the ideal MHD equations, which is the system
that will be solved for the simulations of stellar winds from isolated stars.

Besides the system of Eqs. (\ref{eq_MHD}), we also solve two passive scalar equations:
\begin{equation}
\label{eq_trac}
\begin{aligned}
\frac{\partial \rho s}{\partial t} & + \nabla \cdot \left( \rho s \vec{\upsilon} \right) = 0 \\
\frac{\partial \rho T\!r}{\partial t} & + \nabla \cdot \left( \rho T\!r\, \vec{\upsilon} \right) = 0 \; ,
\end{aligned}
\end{equation}
where $s$ is the specific entropy, whose definition for our newly
implemented equation of state is provided in Appendix \ref{app_eos},
and $T\!r$ is a passive tracer. The entropy is used to monitor the
dissipation and heating usually associated with the numerical
integration of the total energy equation in system of
Eqs. (\ref{eq_MHD}), e.g. by providing a maximum and minimum entropy
value that can be attained during the computation, and it is used to
compute the internal energy when the total energy equation provides an
unphysical value of the latter. The passive scalar $T\!r$ is used to
track the disk material and distinguish it from the coronal/stellar
wind plasma. 

We employ a second-order Godunov method provided by the PLUTO
code\footnote{PLUTO is freely available at
  http://plutocode.ph.unito.it} \citep{Mignone:2007aa} to numerically
solve the system of Eqs. (\ref{eq_MHD})-(\ref{eq_trac}). We use a
mixture of linear and parabolic interpolation to perform the spatial
reconstruction of the primitive variables. 
The approximate HLLD Riemann solver developed by \citet{Miyoshi:2010aa}
is employed to compute the intercell fluxes, exploiting its ability to
subtract the contribution of a potential magnetic field (i.e. the
initial unperturbed stellar magnetosphere) to compute the Lorentz
forces. A second order Runge-Kutta scheme advances the MHD equations in time. 
The hyperbolic divergence cleaning method \citep{Dedner:2002aa} is
used to control the $\nabla \cdot \vec{B} = 0$ condition for the
magnetic field. The viscous and resistive terms, computed using a
second-order finite difference approximation, have been integrated in
time explicitly. We solved the MHD equations in a frame of reference
co-rotating with the star. 

All simulations have been carried out in 2.5 dimensions, i.e. in a
two-dimensional computational domain with three-dimensional vector
fields, assuming axisymmetry around the stellar rotation axis. We
solved the equations in a spherical system of coordinates $(R,
\theta)$. From this point on we will use the capital letter $R$ for
the spherical radius and the lower case $r = R \sin \theta$ to
indicate the cylindrical one. Our computational domain covers a region
$R \in [1,50.76]R_\ast$, where $R_{\ast}$ is the stellar radius, and
$\theta \in [0,\pi]$. We discretized the domain with 320 points in the
radial direction using a logarithmic grid spacing (i.e. $\Delta R
\propto R$) and with 256 points along $\theta$ with a uniform
resolution, so that $R \Delta \theta \approx \Delta R$, i.e. the cells
are approximately square.

\subsection{Initial and boundary conditions}
\label{sec_ic_bc}

In this work we will present simulations of two different systems:
magnetized stars launching thermally-driven winds either (1) isolated
or (2) interacting with an accretion disk. 
For the latter one, 
the initial conditions are made up of three parts: the accretion disk, the
stellar corona, and the stellar magnetic field. 
For the isolated-stellar-wind simulation the initial conditions
are the same, without the presence of a disk.

We set up a Keplerian accretion disk adopting an $\alpha$ parametrization \citep{Shakura:1973aa} for the viscosity.
Neglecting the inertial terms, its thermal pressure $P_\mathrm{d}$ and
density $\rho_\mathrm{d}$ can be determined by the vertical
hydrostatic equilibrium, while the toroidal speed
$\upsilon_{\phi\mathrm{d}}$ can be derived from the radial
equilibrium. Assuming a polytropic condition (i.e., $P_\mathrm{d}
\propto \rho_\mathrm{d}^\gamma$ with $\gamma=5/3$), we obtained
\begin{equation}
\label{eq_disk}
\begin{aligned}
\rho_\mathrm{d} &= \rho_\mathrm{d0} \left \{ \frac{2}{5 \epsilon^2}
\left[\frac{R_{\ast}}{R} - \left( 1 - \frac{5
	\epsilon^2}{2}\right) \frac{R_{\ast}}{r} \right] \right \}^{3/2} \\
P_\mathrm{d} &= \epsilon^2 \rho_\mathrm{d0} \upsilon_{K \ast}^2
  \left(\frac{\rho_d}{\rho_{d0}} \right)^{5/3} \\
\upsilon_{\phi \mathrm{d}} &= \sqrt{\left(1-\frac{5}{2}\epsilon^2\right)\frac{GM_\ast}{r}} \, ,
\end{aligned}
\end{equation}
where $\epsilon = c_\mathrm{sd}/\upsilon_{K}\rvert_{\theta=\pi/2}$ is
the disk aspect ratio given by the ratio between the disk isothermal
sound speed $c_\mathrm{sd}=\sqrt{P_\mathrm{d}/\rho_\mathrm{d}}$ and
the Keplerian speed $\upsilon_K = \sqrt{GM_\ast/r}$ evaluated at the
disk midplane; $\rho_{d0}$ and $\upsilon_{K \ast}$ are the disk
density and Keplerian speed at the disk midplane at $R_{\ast}$. The
accretion speed is computed solving the stationary angular momentum
equation using an $\alpha$ parametrization for the kinematic viscosity
$\nu_\mathrm{v}$: 
\begin{equation}
\label{eq_visc}
\nu_\mathrm{v} = \frac{2}{3}\alpha_\mathrm{v}\frac{c_\mathrm{sd}^2}{\Omega_K} \, ,
\end{equation}
where $\Omega_K = \sqrt{GM_\ast/r^3}$ is the Keplerian angular
speed. Neglecting the $\mathcal{T}_{\theta\phi}$ component of the
stress tensor Eq. (\ref{eq_Vstress}) we obtained 
\begin{equation}
\label{eq_vacc}
  \upsilon_{Rd} = -\alpha_{\mathrm{v}} \frac{c_\mathrm{sd}^2}{\upsilon_K} \sin\theta \, .
\end{equation}
This equation shows that the inertial term due to the accretion flow
in the radial momentum equation is of order
$\mathcal{O}(\alpha_\mathrm{v}^2\epsilon^4)$, while the thermal
pressure gradient is of order $\mathcal{O}(\epsilon^2)$, so that it
was possible to neglect it when deriving the disk equilibrium
Eq. (\ref{eq_disk}). We neglected the $\mathcal{T}_{\theta\phi}$
component of the viscous stress tensor in order to avoid the backflow
along the disk midplane, most likely unphysical, usually associated
with the three-dimensional models of $\alpha$ accretion disks
\citep[see e.g.,][]{Regev:2002aa}. Consistently, we set
$\mathcal{T}_{\theta\phi} = 0$ also during the time-dependent
calculations. 

 We assume that the disk possesses, beside an alpha viscosity,
  also an anomalous magnetic diffusivity allowing the magnetic flux
  not to be perfectly frozen into the accretion flow. The idea behind
  these mechanisms is that some sort of instability, e.g.,
  magneto-rotational or interchange, can trigger a large-scale
  turbulent transport of angular momentum (viscosity) and magnetic
  flux (resistivity). For both viscosity $\nu_\mathrm{v}$ and magnetic
  diffusivity $\nu_\mathrm{m}$ we use a customary alpha
  parametrization: 
\begin{equation}
\label{eq_mdiff}
\begin{aligned}
\nu_\mathrm{v} &= & \frac{2}{3}\alpha_\mathrm{v}\frac{c_{\mathrm{s}\nu}^2}{\Omega_K} \\
\nu_\mathrm{m} &= & \alpha_\mathrm{m}\frac{c_{\mathrm{s}\nu}^2}{\Omega_K} \, ,
\end{aligned}
\end{equation}
akin to Eq. (\ref{eq_visc}), where the isothermal sound speed
$c_{\mathrm{s}\nu}$ is now space and time dependent. In the outer part
of the disk, where the disk structure remains approximately unchanged
with respect to the initial disk structure, we fix the sound speed at
its initial value provided by Eq. (\ref{eq_disk}), while in the inner
part of the disk down to the truncation region, we use the local sound
speed value. We use a function $F(r)$ that smoothly goes from zero for
$r < 0.5 R_\mathrm{t,i}$ to one for $r>1.5 R_\mathrm{t,i}$ (where
$R_\mathrm{t,i}$ is the position at which the initial disk solution is
truncated, see below) to match the two $c_\mathrm{s}$ values. We
assume that, if the magnetic field is strong enough, the instabilities
that trigger the anomalous alpha transport are suppressed. We
therefore multiply the sound speed that defines the transport
coefficients by an exponential function that goes to zero for $\mu =
B^2/8\pi P > 1$. In practice, this term cancels the transport
coefficients outside the disk, while determining their smooth
transition to zero along the accretion funnels. Finally, we multiply
the sound speed by a tracer $T\!r$ that is set to zero in the stellar
atmosphere and to one inside the disk, in order to suppress the
viscosity and resistivity in the stellar wind and the magnetic
cavity. The full expression for $c_\mathrm{s}$ in Eq. (\ref{eq_mdiff})
is therefore: 
\begin{equation}
c_{\mathrm{s}\nu}^2 = \left\{\frac{P}{\rho}\left[1-F(r)\right] +  \frac{P_\mathrm{d}}{\rho_\mathrm{d}}F(r)\right\}
\exp\left\{-\left[\frac{\max\left(\mu,1\right)-1}{2}\right]^2\right\} T\!r \, .
\end{equation}

We initialized the stellar atmosphere surrounding the disk computing
the thermal pressure and density profiles of a one-dimensional,
spherically symmetric, isentropic (according to the entropy $s$
defined in Appendix \ref{app_eos}), transonic Parker-like wind
model. This solution is defined by its density $\rho_\ast$ and sound
speed $c_{s\ast}$ at the stellar surface. The poloidal speed is set to
zero.

The stellar magnetosphere is modeled initially as a potential dipolar 
field aligned with stellar the rotation axis. Its two components are
\begin{equation}
\begin{aligned}
B_R & = 2B_{\ast} \left(\frac{R_{\ast}}{R}\right)^3 \cos\theta\\
B_{\theta} & = B_{\ast} \left(\frac{R_{\ast}}{R}\right)^3 \sin \theta \,,
\end{aligned}
\label{eq_bdip}
\end{equation}
where $B_\ast$ is the magnetic field intensity at the stellar equator.

The interface between the disk surface and the corona is placed at the
position where the disk and coronal thermal pressures are equal. 
We compute the initial truncation radius $R_\mathrm{t,i}$ by solving the
following implicit equation in the variable $R$: 
\begin{equation}
M_s = \left\vert \frac{B_\phi^+ B_\mathrm{d,\theta=\pi/2}}{2\pi P_\mathrm{d,\theta=\pi/2}} \right\vert \, ,
\end{equation}
where $B_\mathrm{d,\theta=\pi/2} = B_\ast(R_\ast/R)^3$ is the
intensity of the initial magnetic field and
$P_\mathrm{d,\theta=\pi/2}$ is the disk thermal pressure both taken at
the disk midplane; $B_\phi^+$ is the toroidal magnetic field at the
disk surface while $M_s$ roughly corresponds to the sonic Mach number
of the accretion flow induced by the large-scale magnetic field torque
\citep[see e.g. Eq. (3) in][]{Combet:2008aa}. For $B_\phi^+$ we take
an estimate of the toroidal field induced at the disk surface by the
star-disk differential rotation \citep[see
e.g.][]{Collier-Cameron:1993aa}: 
\begin{equation}
B_\phi^+ = \frac{B_\mathrm{d,\theta=\pi/2}}{\alpha_m\epsilon}
\left[\left(\frac{R}{R_{co}}\right)^{2/3} -1\right] \, ,
\end{equation}
where $R_{co} = (GM_\ast/\Omega_\ast^2)^{1/3}$ is the corotation
radius, where the disk Keplerian rotation equals the stellar angular
speed $\Omega_\ast$. 
We take $M_s = 1.5$: this means that at its inner radius the disk
starts to accrete trans-sonically, which is a typical condition to
form the accretion funnels \citep[see e.g.][]{Bessolaz:2008aa}, the
initial disk structure Eq. (\ref{eq_disk}) is deeply modified and the
local viscous torque is completely negligible (notice, for example,
that the typical sonic Mach number induced by the viscous torque is of
the order $M_s = \alpha_v \epsilon$, see Eq. \ref{eq_vacc}). Anyway we
set an upper limit for the initial truncation radius $R_{t,i} < 0.8\,
R_{co}$. Finally, the magnetic surfaces initially threading the
disk are set to rotate at the Keplerian angular speed calculated at
the disk anchoring radius, while the magnetic surfaces inside
$R_{t,i}$ are forced to corotate with the star. 

On the rotation axis we assume axisymmetric boundary conditions.
At the $R = R_\ast$ boundary we have to consider two types of
conditions, one for a subsonic inflow (the stellar wind) and another
for a supersonic outflow (the accretion columns). For the subsonic
inflow condition we fix in the ghost zones the density and pressure
profiles of the one-dimensional wind model used to initialize the
stellar atmosphere. For the supersonic outflow conditions the density
and pressure must be left free to adjust to the values in the
accretion funnel: in this region we use a power-law extrapolation
along the magnetic field lines for the density while the pressure is
set assuming a constant entropy value along the magnetic surfaces. For
intermediate situations (i.e. a subsonic outflow or a hydrostatic
corona), we use the sonic Mach number calculated in the first row of
cells of the domain to linearly interpolate between the two boundary
conditions. 
 Notice that, since the sonic Mach number of the flow can change
  with time, this boundary condition allows the stellar areas occupied
  by the accretion spots or the stellar winds to vary in time and
  adjust to evolution of the system.
Clearly, the simulations of stellar winds from isolated
stars employ the subsonic inflow conditions only, i.e. a fixed
pressure and density profile. The boundary conditions for all the
other quantities are the same in both cases. 
The radial component of the magnetic field is kept fixed, so as to
conserve the total stellar flux. The $\theta$ component is left free
to adjust using a linear extrapolation. The polodial speed
$\upsilon_p$ is set to be parallel to the poloidal magnetic field,
using the conservation of the axisymmetric ideal MHD invariant $k =
\rho \upsilon_p/B_p$ along the field lines to determine its
value. This condition ensures a smooth inflow for the stellar wind
injection while the supersonic infalling accretion funnels are
absorbed by the inner boundary without generating a shock. We follow
\citet{Zanni:2009aa,Zanni:2013aa} to impose a boundary condition that
ensures that the magnetic field is frozen into the rotating stellar
surface, that is to say that the poloidal component of the electric
field in the rotating frame of reference must be equal to zero,
i.e. $(\upsilon_\phi -r \Omega_\ast)B_p - \upsilon_p B_\phi = 0$. 
 In order to achieve that, we extract the radial derivative of the
toroidal field $B_\phi$ from the angular momentum conservation equation
 in the system of Eqs. (\ref{eq_MHD}):
\begin{equation}
\label{eq_bphibound}
\frac{\partial B_\phi}{\partial R} = \left[ 
\frac{\rho\vec{\upsilon}}{r}\cdot\nabla\left(r\upsilon_\phi\right)-J_RB_\theta-\frac{B_\phi
  B_R}{4\pi R}+\rho\frac{\partial \upsilon_\phi}{\partial t}
\right] \frac{4\pi}{B_R} \, .
\end{equation}
In order to use this derivative to linearly extrapolate the value of
the toroidal field in the boundary zones, we compute a
finite-difference approximation of the right-hand side of
Eq. (\ref{eq_bphibound}) calculated in the first row of cells in the
computational domain adjacent to the inner boundary. We employ a
first-order approximation for the spatial derivatives while for the
local toroidal acceleration we use the expression  
\begin{equation}
\rho \frac{\partial \upsilon_\phi}{\partial t} = \rho
\frac{r \Omega_{\ast} +  \upsilon_{p} B_{\phi/}/B_{p} -
	\upsilon_\phi}{\Delta t} \, ,
\label{eq_bc_angmom}
\end{equation} 
where $\Delta t$ is the Alfv\'{e}n crossing time of a grid cell at the
inner domain. This condition ensures that the Lorentz force at the
stellar boundary tries to force the magnetic surfaces to rotate at a
rate $\Omega_\ast$ on a timescale $\Delta t$. Consistently, the
boundary value of the toroidal speed is set to 
\begin{equation}
  \upsilon_{\phi} = r \Omega_{\ast} + \upsilon_{p}
  \frac{B_{\phi}}{B_{p}} \, .
  \label{eq_vphistar}
\end{equation}

At the outer radial boundary a power-law extrapolation for
density and pressure is used and all the other variables are linearly
extrapolated. Particular attention is devoted to the boundary
condition for the toroidal field in the region where the stellar wind
exits the computational domain. We used an approach similar to the one
employed at the inner radial boundary, only using a much longer
timescale $\Delta t$ of the order of the Alfv\'{e}n crossing time of
the entire computational domain. This condition avoids artificial
torques exerted on the star even when the matter crosses the outer
boundary at sub-Alfv\'{e}nic speeds. 
 
\subsection{Units and normalization}
\label{sec_units}

Simulations have been performed and results will be presented in dimensionless units.
The stellar radius, $R_{\ast}$ is employed as the unit of
length. Given the stellar mass, $M_{\ast}$, the velocities can be
expressed in units of the Keplerian speed at the stellar radius,
$\upsilon_{K\ast} = \sqrt{GM_{\ast}/R_{\ast}}$. The unit time is
$t_{0} = R_{\ast}/\upsilon_{K\ast}$. Using the density of the stellar
wind at the stellar surface, $\rho_{\ast}$, as the reference density,
magnetic fields are given in units of $B_{0} = \sqrt{4 \pi \rho_{\ast}
  \upsilon_{K\ast}^2}$. Mass-accretion/outflow rates and
torques are expressed in units of $\dot{M}_{0} = \rho_{\ast} R_{\ast}^2
\upsilon_{K\ast}$, and $\tau_{0} = \rho_{\ast} R_{\ast}^3
\upsilon_{K\ast}^2$ respectively. Using $\rho_{\ast} =
10^{-12}\ \mathrm{g}\ \mathrm{cm}^{-3}$,
$R_{\ast} = 2R_{\sun}$, $M_{\ast} = 0.7M_{\sun}$ as reference values, we obtain
\begin{equation}
 \begin{aligned}
   \upsilon_{K \ast} &= 258 \left(\frac{M_{\ast}}{0.7M_{\sun}}
   \right)^{1/2} \left(\frac{R_{\ast}}{2R_{\sun}}  \right)^{-1/2}\ \
   \mathrm{km\ s^{-1}}\\
   t_{0} &= 0.062 \left(\frac{M_{\ast}}{0.7M_{\sun}}\right)^{-1/2}
   \left(\frac{R_{\ast}}{2R_{\sun}}  \right)^{3/2}\ \ \mathrm{days}\\
   B_{0} &= 93.8 \left(\frac{\rho_{\ast}}{\scriptstyle 10^{-12}\ \mathrm{g}\
   	\mathrm{cm}^{-3}}  \right)^{1/2} 
   \left(\frac{M_{\ast}}{0.7M_{\sun}}\right)^{1/2} 
   \left(\frac{R_{\ast}}{2R_{\sun}}  \right)^{-1/2}\ \ \mathrm{G}\\
   \dot{M}_{0} &= 8.32 \times 10^{-9}
   \left(\frac{\rho_{\ast}}{\scriptstyle 10^{-12}\ \mathrm{g}\
   	\mathrm{cm}^{-3}}  \right)
   \left(\frac{M_{\ast}}{0.7M_{\sun}}\right)^{1/2}
   \left(\frac{R_{\ast}}{2R_{\sun}}  \right)^{3/2}\ \
   M_{\sun}\ \mathrm{yr^{-1}}\\
   \tau_{0} &= 1.89 \times 10^{36}
   \left(\frac{\rho_{\ast}}{\scriptstyle 10^{-12}\ \mathrm{g}\
   	\mathrm{cm}^{-3}}  \right)
   \left(\frac{M_{\ast}}{0.7M_{\sun}}\right)
   \left(\frac{R_{\ast}}{2R_{\sun}}  \right)^{2}\ \ \mathrm{dyn\ cm}
   \label{eq_units}
 \end{aligned}  
\end{equation}

\subsection{Parameters of the study}
\label{params}

\begin{table*}[tbp]
  \caption{Varied input parameters and global properties of the simulations.}
  \label{tab_data_sim}
  \centering
    %
    \begin{tabular}{c c c c c c c c c c c}
      \hline\hline
      Case & $B_{\ast}/B_{0}$ & $\Upsilon$ & $\Phi_{open}/\Phi_{\ast}$ & $\Upsilon_{open}$ 
      & $\langle r_A \rangle /R_{\ast}$ & $\dot{M}_{acc}/\dot{M}_{0}$ & $\Upsilon_{acc}$ 
      & $R_t/R_{\ast}$  & $\tau_{sw}/\tau_{acc}$ & $\tau_{me}/\tau_{acc}$ \\
      \hline
      1\tablefootmark{a}  & 0.8125 & 5030    & 0.296  & 441   & 7.98 & ... & ... & ... & ... & ... \\
      2\tablefootmark{a}  & 1.625  & 25800   & 0.230  & 1360  & 12.0 & ... & ... & ... & ... & ... \\
      3\tablefootmark{a}  & 3.25   & 130000  & 0.172  & 3860  & 17.6 & ... & ... & ... & ... & ... \\
      4\tablefootmark{a}  & 6.5    & 635000  & 0.131  & 10900 & 25.6 & ... & ... & ... & ... & ... \\
      5\tablefootmark{a}  & 13     & 2960000 & 0.0951 & 26700 & 33.9 & ... & ... & ... & ... & ... \\
      \hline
      1\tablefootmark{b}  & 1.625 & 45000   & 0.381  & 6550  & 23.4 & 0.489  & 3.82  & 1.82  & -0.355  & 0.486 \\
      2\tablefootmark{b}  & 3.25  & 139000  & 0.267  & 9890  & 28.1 & 0.628  & 11.9  & 2.64  & -0.416  & 0.219 \\
      3\tablefootmark{b}  & 6.5   & 667000  & 0.164  & 18000 & 37.0 & 0.861  & 34.7  & 4.04  & -0.359  & 0.111 \\
      4\tablefootmark{b}  & 9.75  & 2220000 & 0.114  & 28900 & 44.9 & 1.20   & 55.9  & 4.68  & -0.221  & -0.0237 \\
      5\tablefootmark{b}  & 13    & 5420000 & 0.0873 & 41300 & 52.5 & 1.61   & 74.4  & 4.92  & -0.173  & -0.0853\\
      \hline
    \end{tabular}
  \tablefoot{\tablefoottext{a}{Isolated-stellar-wind (ISW)
      simulations.} \tablefoottext{b}{Star-disk-interaction (SDI)
      simulations.} In our simulations, a negative (positive) torque
    indicates angular momentum flowing towards (away from) the
    star (see also \S\ref{sec_sdi}). Consequently, $\tau_{sw}
    > 0$, $\tau_{acc} < 0$, and $\tau_{me}  \lessgtr 0$  (for the
    sign of $\tau_{me}$ in each SDI case see also Fig. \ref{fig_jme_fit}).}
\end{table*}

Two sets of numerical simulations are presented in this work. The
first set includes five simulations of isolated stellar winds
(hereafter ``ISW''), and the second one consists of five simulations of
star-disk-interacting systems (hereafter ``SDI''). 

Once the MHD equations and the initial conditions have been
normalized, the ISW simulations depend on three dimensionless free
parameters, the stellar rotation rate, given as the fraction of the
break-up speed $f_{\ast} = R_\ast \Omega_{\ast}/\upsilon_{K\ast}$, the
magnetic field intensity $B_\ast/B_0$ and the wind sound speed at the
stellar surface $c_{s\ast}/\upsilon_{K\ast}$. The SDI simulations
require four additional parameters to define the disk structure, its
density $\rho_\mathrm{d0}/\rho_\ast$, the aspect ratio $\epsilon$ and
the transport coefficients $\alpha_{v}$ and $\alpha_m$.   
In both sets we fix the stellar rotation taking $f_\ast=0.05$, which
corresponds to a stellar rotation period of 
\begin{equation}
P_{\ast}= 7.83 \left(\frac{M_{\ast}}{0.7M_{\sun}}\right)^{-1/2}
\left(\frac{R_{\ast}}{2R_{\sun}}  \right)^{3/2}\ \ \mathrm{days},
\label{eq_Pstar}
\end{equation}
and a corotation radius $R_{co} = 7.37 \, R_{\ast}$. The stellar wind
sound speed at the stellar surface is assumed to be $c_{s\ast} = 0.35
\, \upsilon_{K\ast}$, which corresponds to a specific enthalpy $h_\ast
= 1.38 \,\upsilon_{K\ast}^2$.   
To define the initial disk structure in the SDI simulations we take in
all cases $\rho_\mathrm{d0} = 100 \, \rho_\ast$, $\epsilon = 0.075$
and $\alpha_v = \alpha_m = 0.2$. With this choice of parameters the
initial disk accretion rate, determined by the viscous torque only, is 
\begin{equation}
\label{eq_vrate}
\dot{M}_{acc,i} \approx 0.12 \left(\frac{\alpha_v}{0.2}\right)
                  \left(\frac{\rho_\mathrm{d0}/\rho_\ast}{100}\right)
                  \left(\frac{\epsilon}{0.075}\right)^3
                  \ \dot{M}_0 \ .
\end{equation}

In the two sets of simulations we varied the stellar magnetic field
strength only. The value of $B_{\ast}/B_{0}$ for all the simulations
presented in this work is listed in the 2nd column of Table \ref{tab_data_sim}. 

\section{Results}

\subsection{MHD Simulations}
\label{sec_mhdsim}

\subsubsection{Numerical Solutions of star-disk interaction}
\label{sec_sdi}

\begin{figure*}
  \centering
  \includegraphics[scale=0.65]{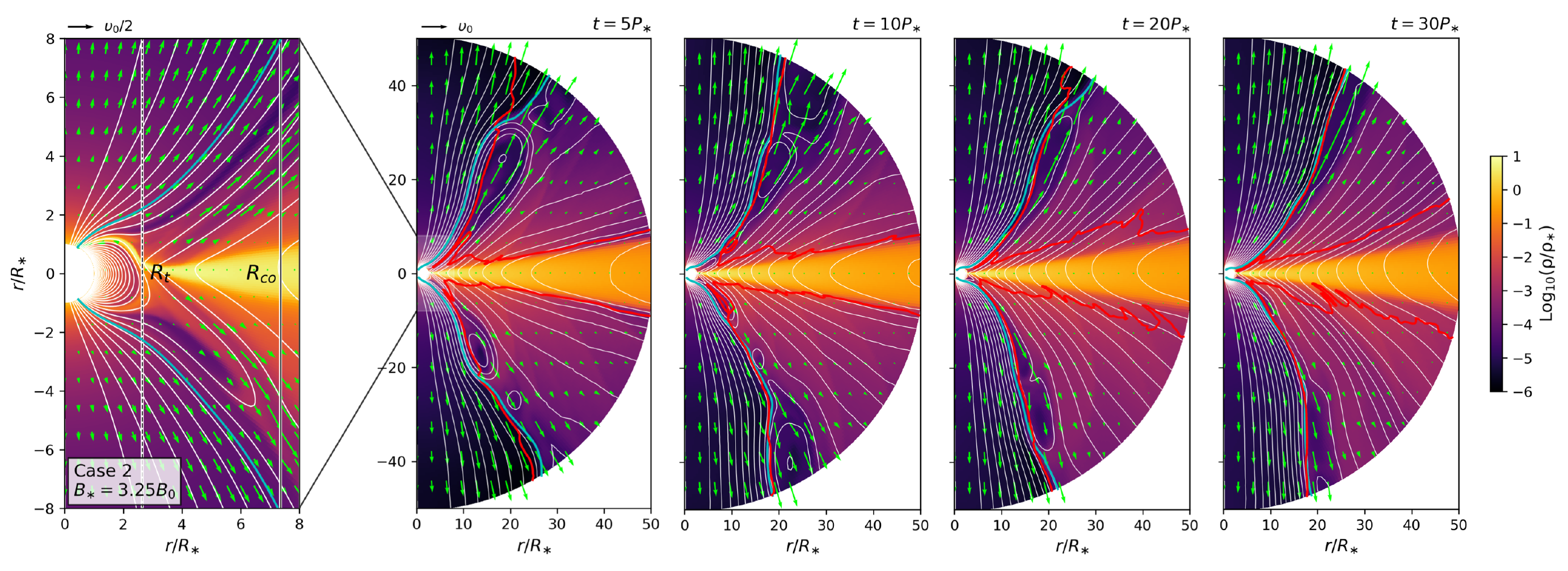}
  \caption{Logarithmic normalized density (colormaps) showing the
      temporal evolution of a star-disk-intercation simulation. The snapshots
      were taken after 5, 10, 20, and 30 stellar rotation periods,
      $P_{\ast}$. The far-left panel illustrates the inner domain of
      this simulation, with the vertical white lines indicating the
      location of the truncation radius $R_t$ (dashed black core) and
      corotation radius $R_{co}$ (solid black core). Each plot
      includes magnetic field lines (white lines) and velocity-field
      vectors. The cyan lines delimit the stellar-wind flux tube and
      the red lines mark the Alfv\'{e}n surface.}
  \label{fig_rho_sdi}
\end{figure*}

We ran all the SDI simulations for up to 30 stellar periods. After a
short strong transient (2-3 stellar periods) the simulations reach a
quasi-stationary state, where the integrated mass and angular momentum
fluxes slowly vary in time around a mean value. All the SDI cases show a
longer-term decrease of the accretion rate, with this effect being
more prominent for higher magnetic field cases. This decrease leads 
the truncation radius getting closer to $R_{co}$ and consequently,
simulations having $B_{\ast} > 3.25 \, B_0$ start to display a strong 
variability most likely associated with a weak propeller regime. We
decided to exclude these later highly variable phases from our
analysis and focus on the steadily accreting stages. However, for all
the SDI cases we based our analysis on timescales longer than 10
stellar periods.

An example of the SDI numerical solutions obtained in this study is
shown in Fig. \ref{fig_rho_sdi}.
Different groups of field lines can be distinguished: 1) polar
magnetic field lines, anchored at the stellar surface that remain open
over the entire simulation; 2) open field lines threading the disk beyond the
corotation radius, with $R_{co} = 7.37 \, R_{\ast}$; 3) closed field
lines threading the disk in the vicinity of the truncation radius
$R_{t} \simeq 4R_{\ast}$, steadily connecting the disk and the 
star, and maintaining approximately their geometry throughout the
duration of the simulation; 4) closed field lines connecting the star and the
disk beyond $R_t$ and within $R_{co}$, periodically evolving through
phases of inflation, reconnection, and contraction; 5) stellar closed
magnetic loops, located below $R_{t}$, forming a dead zone (or
magnetic cavity).

Each group of magnetic field lines is associated with a
different type of plasma flow observed in each SDI solution. A
conical magnetized stellar wind emerges along the open stellar
field lines exerting a braking torque on the star. A disk-wind
flows along the open field lines
attached to the accretion disk, removing angular momentum from the
disk and adding a torque to the viscous one to determine the accretion
rate beyond corotation. Around $R_{t}$, matter is lifted from the disk
to form accretion funnel flows. Through this process the star and the disk
directly exchange angular mometum. Finally,
intermittent ejections propagate within the area neighboring the
stellar- and disk-wind open magnetic surfaces, known as magnetospheric
ejections \citep[hereafter MEs; see e.g.,][]{Zanni:2013aa}. Such outflows 
occur due to the differential rotation between the star and the disk,
which results in the growth of toroidal field pressure. This process
leads to the inflation of the field lines attached close to $R_{t}$ that
eventually will reconnect, producing plasmoids that propagate
ballistically outwards.  
 Clearly, since they occur in a low plasma $\beta = 8\pi P/B^2$
  region, where $\nu_\mathrm{m} = 0$, these reconnection events are
  numerically driven and therefore depend both on the grid resolution
  and the numerical algorithm employed. On one hand we are rather
  confident that numerical effects do not have a strong impact on the
  launching mechanism of the MEs since, from the point of view of the
  disk, MEs are launched as magneto-centrifugal flows for which
  reconnection is not an acceleration driver. On the other hand,
  magnetic reconnection can modify some large-scale properties of the
  MEs. For example, numerically-driven magnetic reconnection can modify the
stellar magnetic torque due to MEs, since the star can exchange
angular momentum only with parts of the system to which it is
magnetically connected and once the plasmoids disconnect they can not
contribute to the stellar torque anymore. Numerical dissipation can
also modify the position of the reconnection X-point and the
periodicity of the reconnection events, which should be proportional
to the beating frequency (i.e. the difference between the stellar and
disk rotation frequency at $R_t$). Clearly, MEs are the part of our
solutions which can be more sensible to numerical effects. While the
MEs dynamical picture is physically sound, as confirmed by the number
of independent studies presenting an analogous phenomenology
\citep[see e.g.,][]{Hayashi:1996aa,Romanova:2009ab}, some quantitative
aspects such as the mass and angular momentum stellar fluxes
associated with them must be taken with caution.

The main objective of this work is to quantify and model the
contribution of all these type of flows to the stellar angular momentum.
Therefore, we compute their global
properties, in particular their mass flux $\dot{M}$, and angular momentum flux
$\tau$, using, respectively
\begin{equation}
  \dot{M} = \int_{S} \rho \vec{\upsilon_p} \cdot d\vec{S}
  \label{eq_mdot}
\end{equation}

\begin{eqnarray}
  \tau = \int_{S} r \left( \rho \upsilon_{\phi}
    \vec{\upsilon_p} - \frac{B_{\phi} \vec{B_p}}{4 \pi}\right) \cdot d\vec{S} 
  \label{eq_jdot}
\end{eqnarray}
For the SDI simulations, these integrals are calculated at the stellar surface,
separating it into the areas corresponding to the different flow
components, using different subscripts for stellar winds ($sw$),
magnetospheric ejections ($me$) and accretion ($acc$). 
 We ensure that the sum of the mass and angular momentum fluxes
  computed for each flow component corresponds to the total flux
  crossing the inner boundary of our domain. It should be noted that
  the integrals computed inside the magnetic cavity do not contribute
  to the mass and angular momentum fluxes, at least in a time-averaged
  sense. We adopt a sign convention for integrals
  Eqs. (\ref{eq_mdot}), (\ref{eq_jdot}) so that a positive (negative) 
value of $\dot{M}$, $\tau$ denotes mass/angular momentum flowing away
from (towards) the star. Finally, the unsigned magnetic flux is defined as:
\begin{equation}
  \Phi = \int_{S} \vert \vec{B} \cdot d\vec{S} \vert.
  \label{eq_phi}
\end{equation} 
This quantity can be defined for the full stellar surface
($\Phi_\ast$) or for the open flux carried by the stellar wind only
($\Phi_{open}$).

\subsubsection{Numerical solutions of isolated stellar winds}
\label{sec_isw}

\begin{figure}
  \centering
  \includegraphics[scale=0.5]{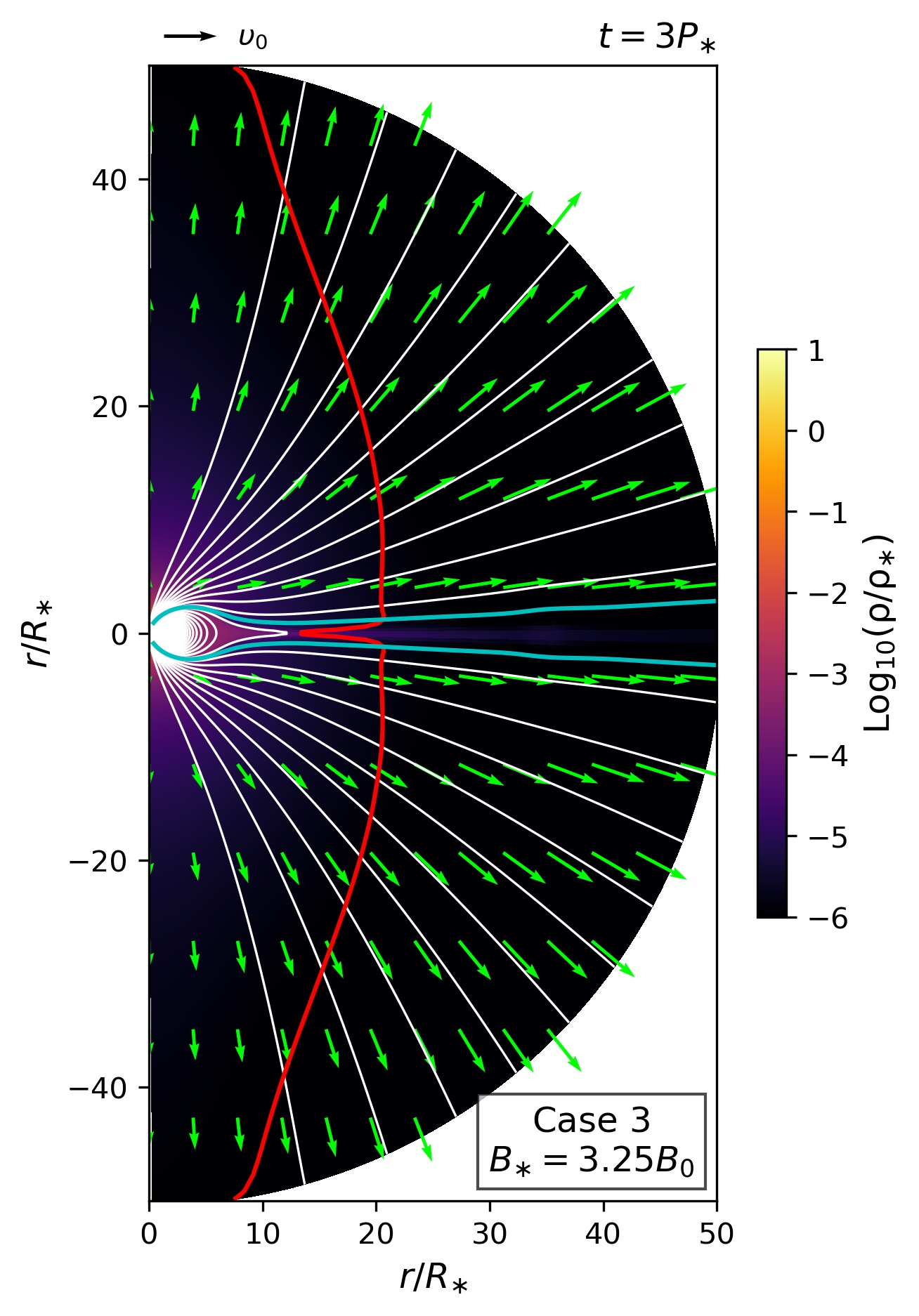}
  \caption{Same as the right panel of Fig. \ref{fig_rho_sdi} for a
    simulation of an isolated stellar wind. The image was taken at 3
    stellar periods.}
  \label{fig_rho_isw}
\end{figure}

Each ISW simulation is stopped when it becomes
quasi-stationary. For this set of simulations, a steady state is
achieved after 2-3 stellar periods. In Fig. \ref{fig_rho_isw},
we present an example of the ISW solutions obtained in this work. Such
simulations are less dynamic, compared to the SDI solutions shown
above, and only two regions can be identified in the plot: a
stellar wind and a dead zone. Clearly, there is a main difference
between the two stellar-wind solutions from the two different
systems (i.e., ISW vs. SDI) studied here. As it can be seen in Fig.
\ref{fig_rho_isw}, asymptotically the flow entirely opens the stellar
magnetosphere, filling with plasma the whole domain. On the other
hand, the SDI stellar-wind solution, illustrated in Fig. \ref{fig_rho_sdi}, is
confined within a conical flux tube due to the presence of
MEs. As we will show later, this difference in the expansion of
the two stellar outflows has a significant effect on their
magnetic torque efficiency. 

Analogously to SDI simulations, Eqs. (\ref{eq_mdot}) and
(\ref{eq_jdot}) are used to determine the mass and angular momentum
fluxes of the stellar wind, and Eq. (\ref{eq_phi}) to compute both the
open and total magnetic flux. Since the ISW simulations converge to a
steady state, these integrals are also averaged in time and space,
using spherical sections of the open flux tubes at different radii, in
order to reduce noise and errors. 

We recall here that Eq. (\ref{eq_jdot}) can be rewritten as
\begin{equation}
  \tau_{sw} = \int_{S} \Lambda \rho \vec{\upsilon_p} \cdot d\vec{S},
  \label{eq_jdot_isw}
\end{equation}
where $\Lambda$ is the total specific angular momentum
carried away by the stellar wind,
\begin{equation}
  \Lambda = r \left(\upsilon_{\phi} - B_{\phi} \frac{B_p}{4
      \pi \rho \upsilon_p} \right).
  \label{eq_lambda}
\end{equation}
For axisymmetric, ideal MHD, steady-state flows, $\Lambda$ is
invariant along magnetic surfaces and, 
in the case of a trans-Alfv\'{e}nic flow, is equal to
\begin{equation} 
\Lambda = \Omega_{\ast} r_{A}^2 ,
\label{eq_ra} 
\end{equation}
where $r_A$ is the wind Alfv\'{e}n radius, the radial distance at
which the stellar outflow reaches the local Alfv\'{e}n velocity
\citep[e.g.,][]{Weber:1967aa,Mestel:1968aa,Mestel:1999aa}. Note that,
$r_A$ represents the distance from the rotation axis or, in other
words, the cylindrical Alfv\'{e}n radius, $r_A = R_A\mathrm{sin}\
\theta_A$, where $R_A$ is now the spherical Alfv\'{e}n
radius and $\theta_A$ the angular distance of the Alfv\'{e}n point
from the rotation axis. The term $r_A/R_{\ast}$ defines a
dimensionless lever-arm that determines the efficiency of the
braking torque acting on the star.

\subsection{Torques scaling}
\label{sec_scallaws}

\subsubsection{Magnetospheric accretion torque}
\label{sec_acc_torq}

The truncation, or magnetospheric, radius $R_t$ of a
star-disk-interacting system is commonly parametrized as
\citep[see e.g.,][]{Bessolaz:2008aa,Zanni:2009aa,Kulkarni:2013aa}, 
\begin{equation}
R_t = K_A R_A,
\label{eq_rt_ra}
\end{equation}
where $R_A$ is the Alfv\'{e}n radius of a spherical free-fall collapse
\citep[e.g.,][]{Lamb:1973aa,Elsner:1977aa}, given as
\begin{equation}
R_A = \left( \frac{B_{\ast}^4 R_{\ast}^{12}}{2 G
	M_{\ast}\dot{M}_{acc}^2} \right)^{1/7} \, ,
\label{eq_ra_ff}
\end{equation}
and $K_A$ is a dimensionless constant that parametrizes the different
geometry and dynamics of disk accretion compared to a free-fall
collapse. For example, \citet{Bessolaz:2008aa,Zanni:2009aa} showed
that $K_A$ can be expressed as a function of the $\beta = 8\pi P/B^2$
parameter and the sonic Mach number of the accretion flow in the
truncation region. Nevertheless, $K_A$ turns out to be always of order
unity \citep[see e.g.,][]{Kulkarni:2013aa,Zanni:2013aa}.

In the present work, we extract the position of the truncation radius
from the simulations by taking the first closed magnetic surface that
envelopes the accretion funnels and look for its intersection with the
accretion disk. This is achieved by searching for the radial position
of the entropy minimum along this magnetic field line, since the bulk
of the disk is characterized by the minimum entropy in the whole
domain. This criterion provides the radius at which the accretion flow
of the disk starts to be deviated and uplifted to form the accretion
columns (see Fig. \ref{fig_rho_sdi}). Another common approach
\citep[see e.g.,][]{Romanova:2002aa,Kulkarni:2013aa} is to look at the
position where the magnetic energy equals the total (thermal plus
kinetic) energy of the disk. This definition provides the position at
which the midplane accretion flow of the disk is completely disrupted
and reasonably provides a slightly smaller estimate of the truncation
radius compared to our method. 

As discussed in Sect.
\ref{sec_sdi}, in the five SDI simulation presented here,
$R_t/R_{\ast}$ slowly increases with time, with variations of $\sim 10\%$
between the lowest and highest value. The time-averaged
$R_t/R_{\ast}$, for all the cases of this work, is given in the 9th
column of Table \ref{tab_data_sim}.

\begin{figure}
	\centering
	\resizebox{\hsize}{!}{\includegraphics{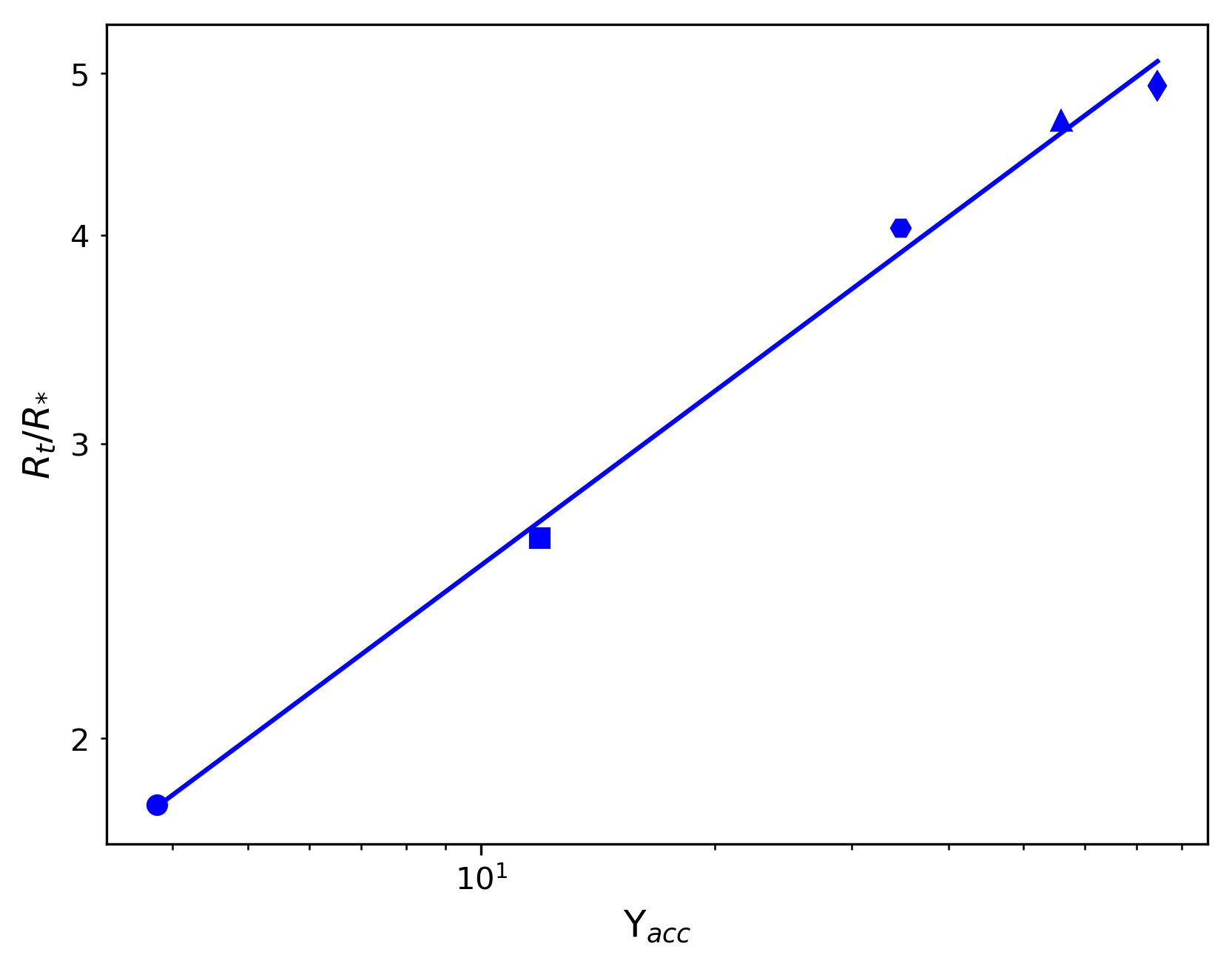}}
	\caption{Normalized truncation radius, $R_t/R_{\ast}$, as a function
		of parameter $\Upsilon_{acc}$. Symbols are the same as in Fig.
		\ref{fig_ra_y}. The solid line shows the fitting function
		(\ref{eq_rt_fit_func}), with $K_t =  1.1452 $ and $m_t = 0.35$.}
	\label{fig_rt_fit}
\end{figure}
Defining an $\Upsilon$-like parameter for the accreting flow,
\begin{equation}
\Upsilon_{acc} = \frac{B_{\ast}^2 R_{\ast}^2}{4 \pi \dot{M}_{acc}
	\upsilon_{esc}} \, ,
\label{eq_yacc}
\end{equation}
equation (\ref{eq_rt_ra}) can now be rewritten as
\begin{equation}
\frac{R_t}{R_{\ast}} = K_A (4 \pi)^{2/7} \Upsilon_{acc}^{2/7}.
\label{eq_rt_yacc}
\end{equation}
 The absolute values of the mass accretion rates and $\Upsilon_{acc}$
 for all the simulations of this study are  
listed in the 7th and 8th column of Table \ref{tab_data_sim}. The
dependence of $R_t/R_{\ast}$ on $\Upsilon_{acc}$ is shown in
Fig. \ref{fig_rt_fit}. We fit the data with a power-law in the form of
\begin{equation}
\frac{R_t}{R_{\ast}} = K_t \Upsilon_{acc}^{m_t} \, ,
\label{eq_rt_fit_func}
\end{equation}
where $K_t, m_t$ are dimensionless fitting constants. This expression
reduces to Eq. (\ref{eq_rt_yacc}) for $m_t = 2/7$. 
The values of $K_t$ and $m_t$, for the best fit (solid line) in the
plot, are $K_{t} = 1.1452$ and $m_{t} = 0.35$, also given in Table
\ref{tab_fit_sdi}. From Eq. (\ref{eq_rt_yacc}), we get $K_A \approx
0.56$, which is in agreement with previous numerical studies
\citep[][]{Long:2005aa,Zanni:2009aa,Zanni:2013aa,Kulkarni:2013aa}. The
power index, $m_t$, is found to be 20\% higher than its theoretical
value, $m_t^{th} = 2/7 \approx 0.286$. This could be the consequence of the
magnetosphere being compressed by the accreting matter, which leads to 
a slower decline of $B_{\ast}$ with $R$ \citep[see e.g., Fig. 8
in][]{Zanni:2009aa}. In addition, $m_t$ differs by almost a factor of
two compared with the value of $m_t = 0.2$ reported in
\citet{Kulkarni:2013aa}. However, their simulations focused on
non-axisymmteric dipolar fields and perhaps further investigation on
the dependence of $K_t$ and  $m_t$ on the inclination of a stellar
magnetosphere is required.

For a Keplerian disk, the specific angular momentum at $R_t$ is $l =
\sqrt{GM_\ast R_{t}}$. Then, the magnitude of the accretion spin-up
torque, as angular momentum is transferred to the star, can be written
as
\begin{equation}
\tau_{acc} = -K_{acc}\dot{M}_{acc} (GM_{\ast}R_t)^{1/2},
\label{eq_torque_acc_fit_func}
\end{equation}
where $K_{acc}$ is a dimensionless constant whose value is related to the
disk rotational profile in the truncation region. The negative sign
indicates that the angular momentum is carried towards the stellar
surface (see also \S\ref{sec_sdi}). Different studies
\citep[][]{Long:2005aa,Kluzniak:2007aa,Zanni:2009aa,Zanni:2013aa}
have shown that the disk is likely to become sub-Keplerian below the 
corotation radius. There are two processes torquing down the
disk in this region. First, the small annulus of the disk that is threaded
by the stellar magnetosphere and corresponds to the base of the
accretion columns is subject to a magnetic-braking torque by the
stellar rotation trying to force the matter to corotate with the star.  
Through this
mechanism, the disk directly transfers its angular momentum to the
star so that, even if this process can modify the Keplerian rotation
profile, it should not have a strong effect on the value of $K_{acc}$.
Second, the region of the disk which corresponds to the base 
of the MEs is subject to an external torque since
MEs can directly remove angular momentum from the surface of the accretion
disk. If this torque is strong enough, it can yield a sub-Keplerian
rotation, with the excess angular momentum being ejected, instead of
being transfered to the star. Therefore this process usually
determines a value of $K_{acc}$ less than unity. 

At this point it should be noted that the torque provided by these mechanisms
can explain the increase of $\dot{M}_{acc}$ with $B_\ast$
observed in our simulations. The magnetic torques are the
dominant drivers of mass accretion in the star-disk-interaction
region, as confirmed by the fact that the mass accretion rates
measured in our models (see Table \ref{tab_data_sim}) are sensibly
larger than the accretion rate determined by the viscous torque only,
see Eq. (\ref{eq_vrate}). As a consequence, a higher $B_{\ast}$ leads
to a stronger overall magnetic torque acting on the disk, which in
turn leads to a higher accretion rate. On the other hand, it is not guaranteed
that the torques driving accretion in the truncation region are
matched by the accretion drivers in the outer disk (viscous and
disk-wind torques). Therefore, the long-term decrease of the accretion
rate and the corresponding increase of the truncation radius, observed
in many of our simulations, is likely due to this mismatch.

\begin{figure}
	\centering
	\resizebox{\hsize}{!}{\includegraphics{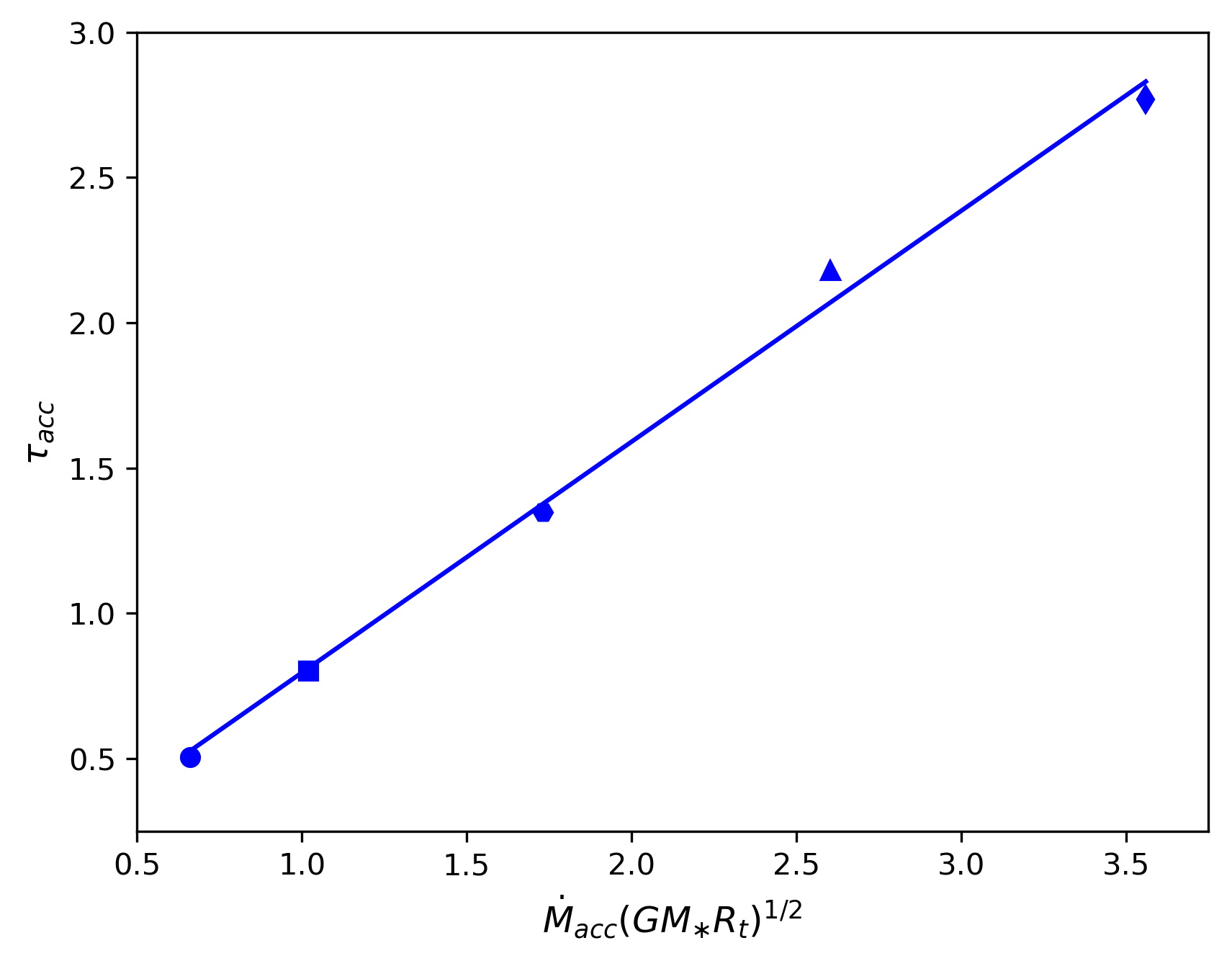}}
	\caption{Normalized accretion torque (absolute values),
		$\tau_{acc}$, versus quantity $\dot{M}_{acc}
		(GM_{\ast}R_t)^{1/2}$. Symbols are the same as in Fig.
		\ref{fig_ra_y}. The solid line corresponds to the best fit of the data, 
		using Eq. (\ref{eq_torque_acc_fit_func}) with $K_{acc} = 0.79$.}
	\label{fig_jacc_fit}
\end{figure}

The absolute value of the accretion torque $\tau_{acc}$ versus the 
quantity $\dot{M}_{acc} (GM_{\ast}R_t)^{1/2}$ is plotted in Fig.
\ref{fig_jacc_fit}. In the figure, the best fit (solid line) provides
	$K_{acc} =  0.79$ (see also Table \ref{tab_fit_sdi}).

By combining Eqs. (\ref{eq_rt_fit_func}) and
(\ref{eq_torque_acc_fit_func}), we obtain the accretion torque
formulation
\begin{equation}
\tau_{acc} = - K_{acc} K_{t}^{1/2} (4 \pi)^{-m_t/2} \sqrt{0.5}\
\upsilon_{esc} ^{(2-m_t)/2}\ R_{\ast}^{m_t}\
\dot{M}_{acc}^{(2-m_t)/2}\ B_{\ast}^{m_t} \, .
\label{eq_torque_acc}
\end{equation}

\subsubsection{Stellar-wind torque}
\label{sec_sw_torq}

\begin{figure}
  \centering
  \resizebox{\hsize}{!}{\includegraphics{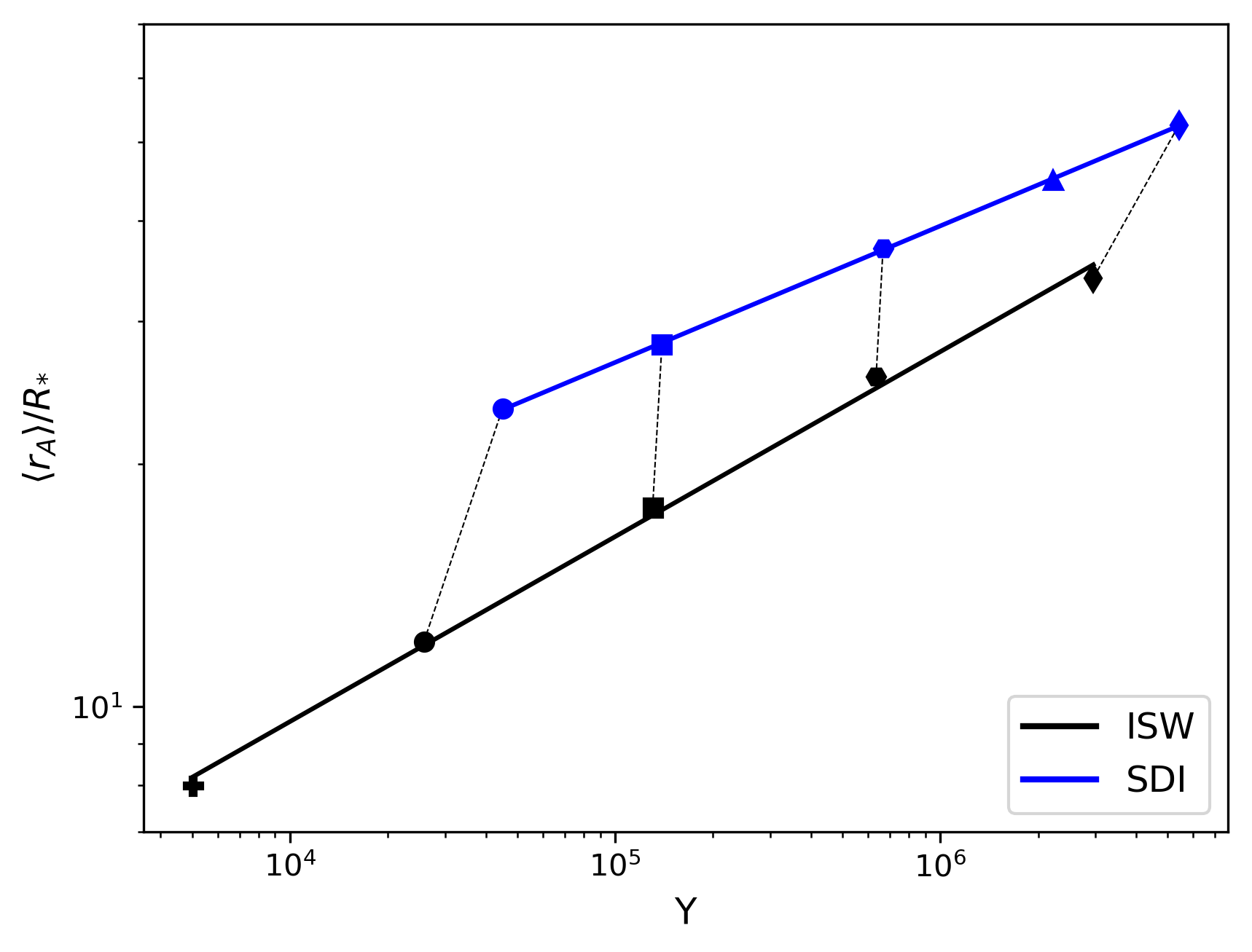}}
  \caption{Normalized effective lever arm, $\langle r_A
    \rangle/R_{\ast}$ as a function of the wind magnetization,
    $\Upsilon$ for all the cases of this work. Each data point
    in the plot represents a single simulation. The blue and black 
    colored points/fitting curves correspond to star-disk-interaction
    (SDI) and isolated-stellar-wind simulations
    (ISW) respectively. Data points with the same
    symbol, connected with a black dashed lines, correspond to numerical
    solutions having the same surface magnetic field strength,
    $B_{\ast}$.}
  \label{fig_ra_y}
\end{figure}

\begin{figure}
  \centering
  \resizebox{\hsize}{!}{\includegraphics{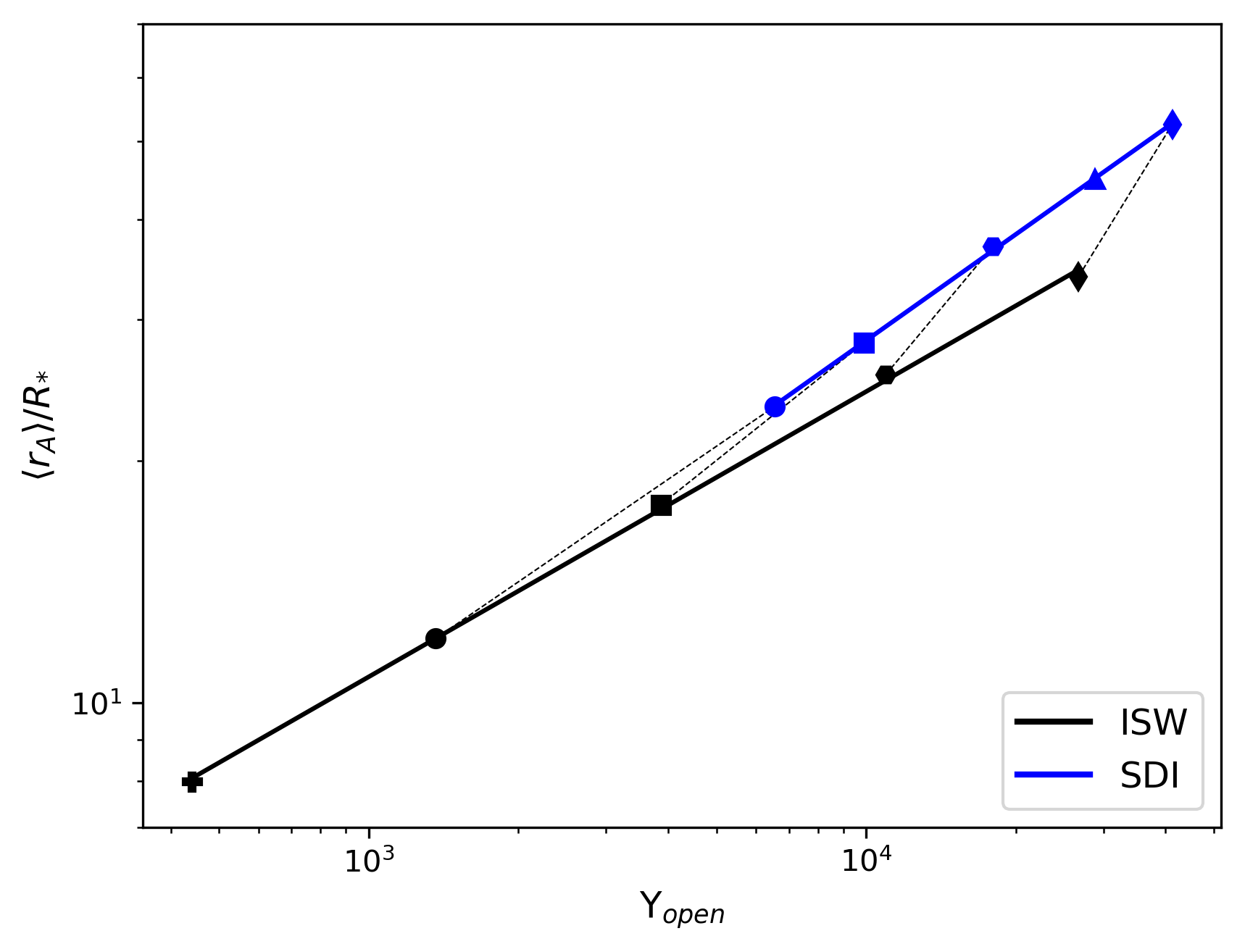}}
  \caption{$\langle r_A \rangle/R_{\ast}$ versus parameter
    $\Upsilon_{open}$ for all the simulation of this study. Colours,
    symbols, and line styles have the same meaning as in Fig.
    \ref{fig_ra_y}.} 
  \label{fig_ra_yopen}
\end{figure}

\begin{figure}
  \centering
  \resizebox{\hsize}{!}{\includegraphics[scale=0.5]{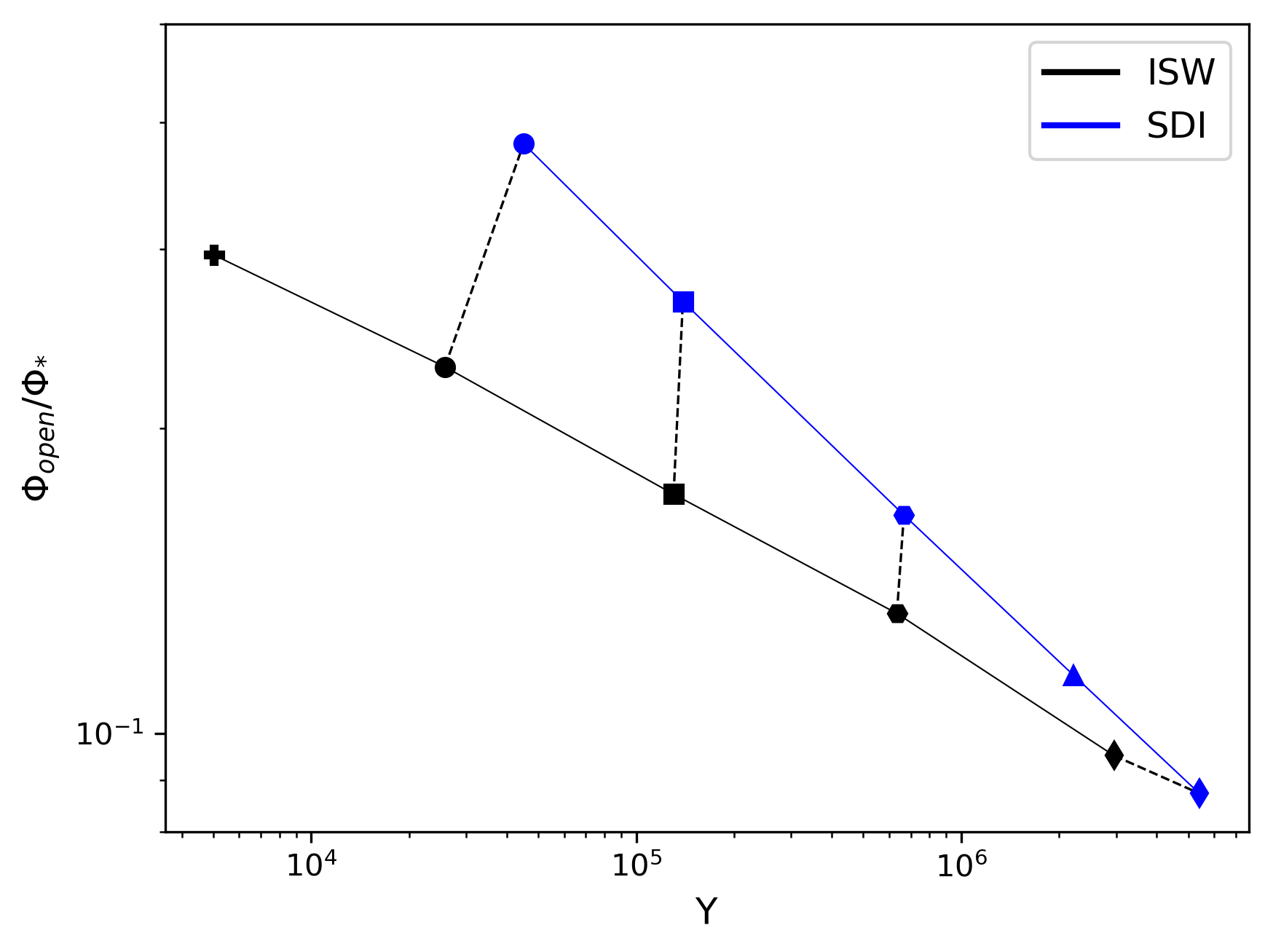}}
  \caption{Fractional open flux $\Phi_{open}/\Phi_{\ast}$ versus
    $\Upsilon$. Colours, symbols, and line  styles have the same
    meaning as in Fig. \ref{fig_ra_y}.}
\label{fig_opflux_y}
\end{figure}

\begin{figure*}[h]
  \centering
  \begin{tabular}{ll}
    \includegraphics[scale=0.5]{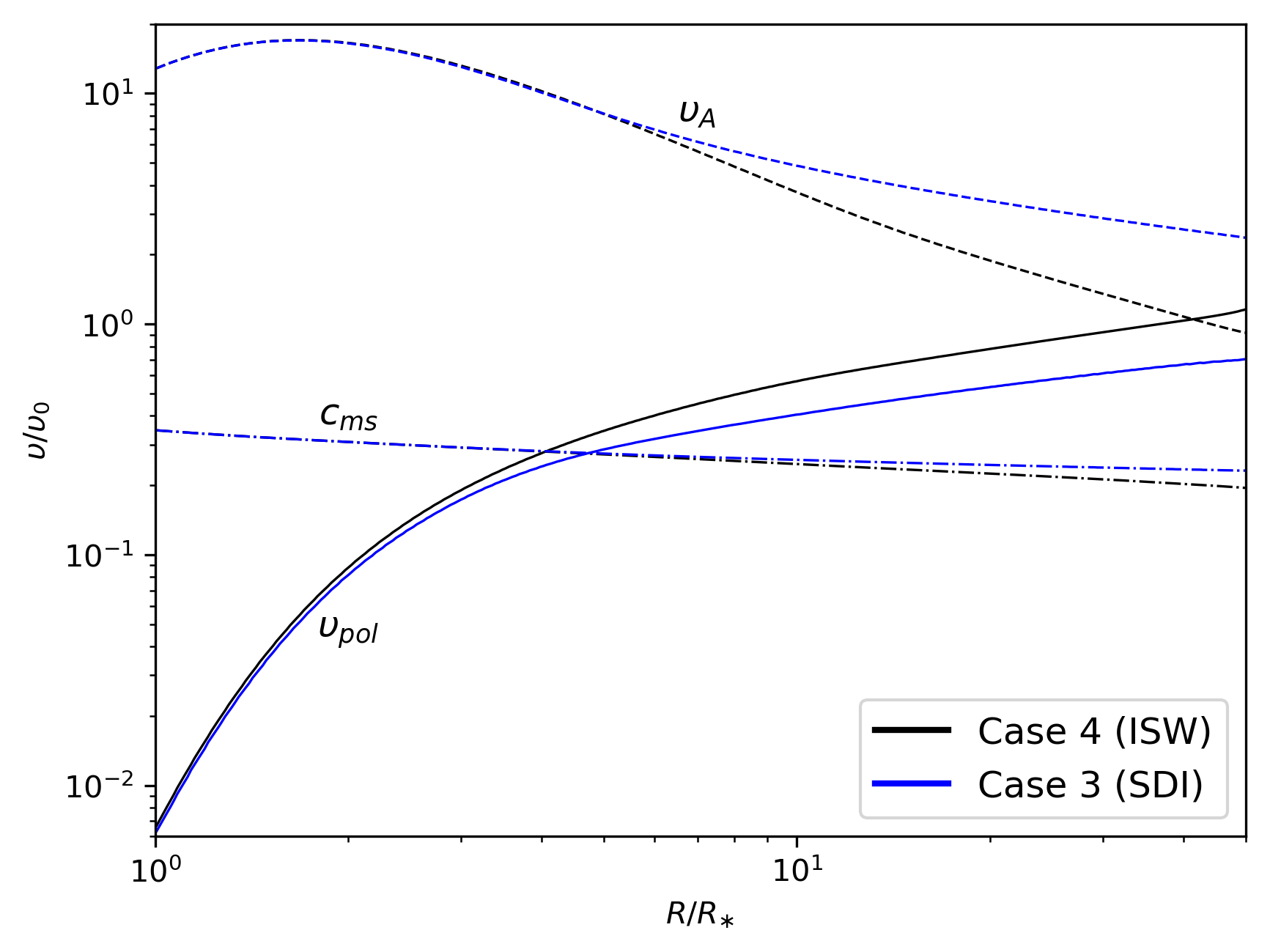}
    &
    \includegraphics[scale=0.5]{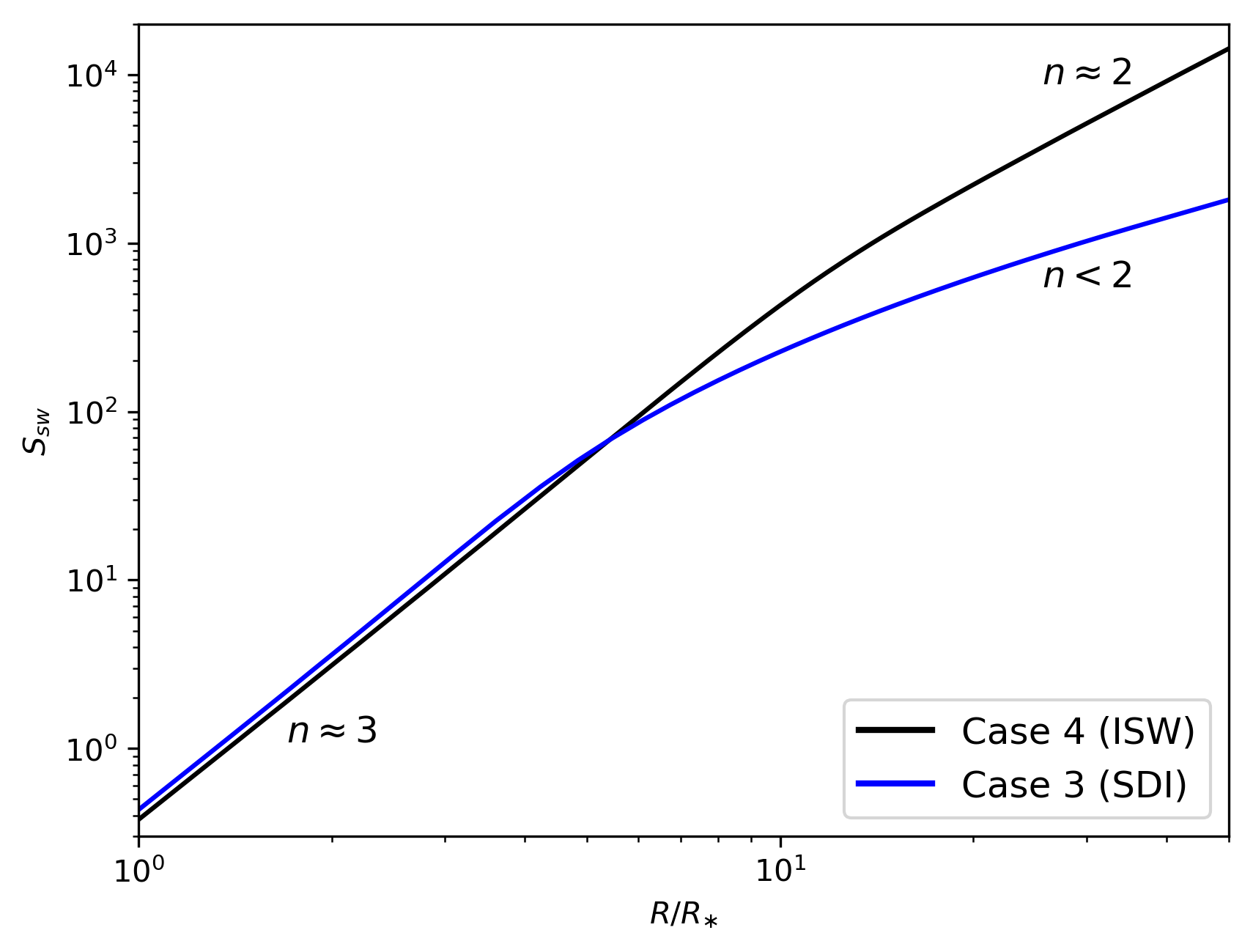}
  \end{tabular}
\caption{(Left) Profiles of normalized Alfv\'{e}n (dashed), slow-magnetosonic
  (dash-dotted), and stellar-wind-poloidal (solid) speeds as a function
  of radius, for two cases presented in this work. The velocities are
  averaged over the $\theta$ coordinate and in time. (Right) Surface
  area of the stellar-wind flux tube, in normalized units, versus
  $R/R_{\ast}$. $S_{sw}$ scales with radial distance as $S_{sw}
  \propto R^n$, where $n \approx 3$ indicates super-radial expansion,
  $n \approx 2$, and $n< 2$ indicate radial and sub-radial expansion,
  respectively. In both panels the colors are the same as in Fig.
  \ref{fig_ra_y}.}
\label{fig_vprof_area}
\end{figure*}

\begin{table}
  \caption{Best-fit coefficients to Eqs. (\ref{eq_ra_upsilon}) and
      (\ref{eq_ra_upsilon_open}).}
  \label{tab_fit_sw}
  \centering
  \begin{tabular}{c c c}
    \hline\hline
    & ISW & SDI \\ 
    \hline
    $K_{sw,s}$ & $1.163\pm0.003$ & $3.8156 \pm 0.0001$\\
    $m_{s}$ & $ 0.229\pm0.007$ & $0.169 \pm 0.001$\\
    $K_{sw,o}$ & $0.9275\pm0.0009$ & $ 0.495\pm0.001$ \\
    $m_{o}$ & $0.355\pm0.005$ & $0.439\pm0.005$\\
    \hline
  \end{tabular}
\end{table}

Combining Eqs. (\ref{eq_mdot}), (\ref{eq_jdot_isw}) and (\ref{eq_ra}),
the torque due to a magnetized stellar wind \citep[see 
e.g.,][]{Schatzman:1962aa,Weber:1967aa,Mestel:1968aa,Mestel:1999aa}
can be written as 
\begin{equation}
  \tau_{sw} = \dot{M}_{sw} \Omega_{\ast} R_{\ast}^{2} {\left( \langle
      r_{A} \rangle \over R_{\ast} \right)^{2}} \, ,
  \label{eq_torque_sw}
\end{equation}
where the positive sign in Eq. (\ref{eq_torque_sw}) denotes
angular momentum carried away from the star (see also \S\ref{sec_sdi}).
In multidimensional solutions the flow becomes super-Alfv\'{e}nic
along a surface (see e.g., Fig. \ref{fig_rho_isw}) and therefore, in
Eq. (\ref{eq_torque_sw}), $\langle r_A \rangle$ represents the wind average lever arm \citep[see,
e.g.,][]{Washimi:1993aa,Matt:2012ab,Reville:2015ab,Pantolmos:2017aa,Finley:2018aa}. Actually
Eq. (\ref{eq_torque_sw}), provides the definition of the effective 
lever arm $\langle r_A \rangle$, which is calculated using 
\begin{equation}
  \frac{\langle r_{A} \rangle}{R_{\ast}} = \left(
    \frac{\tau_{sw}}{\dot{M}_{sw} \Omega_{\ast} R_{\ast}^{2}}
    \right)^{1/2} \, .
  \label{eq_ra_sw}
\end{equation}
The 6th column of Table \ref{tab_data_sim} lists
all the values of $\langle r_A\rangle/R_{\ast}$ for this study.

Following \citet{Matt:2008aa}, we look for a scaling of the lever arm
$\langle r_A\rangle$ with the wind magnetization, $\Upsilon$, defined
as  
\begin{equation} 
  \Upsilon = \frac{\Phi_{\ast}^2}{4 \pi R_{\ast}^2 \dot{M}_{sw}
    \upsilon_{esc}},
  \label{eq_upsilon}
\end{equation}
where $\Phi_{\ast}$ is the total unsigned surface magnetic flux and
$\upsilon_{esc}$ is the escape speed from the surface of the
star. The parameter $\Upsilon_{open}$, which depends on the open unsigned
magnetic flux $\Phi_{open}$ carried by the stellar wind, \citep[see e.g.,][]{Reville:2015ab}, is defined as 
\begin{equation}
  \Upsilon_{open} =\frac{\Phi_{open}^2}{4 \pi R_{\ast}^2
    \dot{M}_{sw} \upsilon_{esc}}.
  \label{eq_upsilon_open}
\end{equation}
For all the simulations of this study, parameters $\Upsilon$ and
$\Upsilon_{open}$ are tabulated in the 3rd and 5th column of
Table \ref{tab_data_sim}, respectively.

As explained in greater detail in Appendix \ref{app_sw_torq},
simple power-laws are usually expected and employed to parametrize
the relation between $\langle r_A \rangle/R_{\ast}$ and $\Upsilon$
or $\Upsilon_{open}$. 
The dependence of $\langle r_A \rangle/R_{\ast}$ on the wind
magnetization, $\Upsilon$, is shown in Fig. \ref{fig_ra_y}. The
following function is used to fit the points
\begin{equation}
 \frac {\langle r_{A}\rangle}{R_{\ast}} = K_{sw,s} \Upsilon^{m_s},
  \label{eq_ra_upsilon}
\end{equation}
where $K_{sw,s}$, $m_{sw,s}$ are dimensionless fitting constants. Two scaling
laws are shown in the plot, which correspond to the two different sets
of simulation studied here (i.e., ISW and SDI simulations). The values 
of $K_{sw,s}$, $m_{sw,s}$ are given in Table \ref{tab_fit_sw}
\footnote{\citet{Matt:2008aa} defined the wind magnetization as
  $\Upsilon = B_{\ast}^2 R_{\ast}^2/(\dot{M}_{sw} \upsilon_{esc})$. To
  compare our $K_{sw,s}$ value with those reported in previous studies
  \citep{Matt:2008aa,Matt:2012ab,Reville:2015ab,Pantolmos:2017aa,Finley:2017aa,Finley:2018aa},
  one should multiply $K_{sw,s}$ by $(4 \pi)^{m_s}$.}.
Figure \ref{fig_ra_y} shows that for a given value of
$\Upsilon$, a stellar wind originating from a star-disk-interacting
system has a larger braking lever arm. 

The dependence of $\langle r_{A}\rangle/R_{\ast}$ on
$\Upsilon_{open}$, for all the cases of the study, is presented in
Fig. \ref{fig_ra_yopen}. The data points are fitted with a power-law
function
\begin{equation}
  \frac {\langle r_{A}\rangle}{R_{\ast}} = K_{sw,o} \Upsilon_{open}^{m_o},
  \label{eq_ra_upsilon_open}
\end{equation}
where $K_{sw,o}$, $m_{sw,o}$ are dimensionless fitting constants. 
The values of $K_{sw,o}$, $m_{sw,o}$ of the two
fits are tablulated in Table \ref{tab_fit_sw}
\footnote{\citet{Reville:2015ab} defined $\Upsilon_{open}$ without a
  $4 \pi$ at the denominator. In order to compare the $K_{sw,o}$
  values with the \citet{Reville:2015ab} results, \citep[see
  also][]{Reville:2016aa,Pantolmos:2017aa,Finley:2017aa,Finley:2018aa},
  one should divide $K_{sw,o}$ by a factor $(4 \pi)^{m_o}$.}.
For clarity, the fitting constants derived from the SDI
  simulations are also listed in Table \ref{tab_fit_sdi}.
As in Fig. \ref{fig_ra_y}, for a given
$\Upsilon_{open}$ value the stellar wind in a star-disk-interacting
system provides a larger magnetic lever arm, even if in this case the
offset between the two curves is clearly smaller. 

We try now to understand the reason behind the SDI and ISW
differences. We verified that the presence of an accretion disk, the
accretion columns and the MEs modify the properties of an isolated
stellar wind in two main ways.

On one hand we observed that, in the SDI simulations, the amount of
open flux that can be exploited by the stellar winds can change
substantially. In Fig. \ref{fig_opflux_y}, we plot the fractional open flux,
given in the 4th column of Table \ref{tab_data_sim}, versus parameter
$\Upsilon$, for all the numerical solutions. In the plot, we identify
two trends. First, for both sets of simulations,
$\Phi_{open}/\Phi_{\ast}$ decreases as a function of $\Upsilon$ and
second, for a given value of $\Upsilon$, SDI simulations tend to
produce a larger amount of fractional open flux.
In the case of ISWs the decrease of the fractional open flux with
$\Upsilon$ can be understood if we look at the $\Upsilon$ parameter as
the ratio between the magnetic field energy density and the kinetic energy density of
the flow \citep[see e.g.,][]{Ud-Doula:2002aa,Ud-Doula:2009aa}. The
larger the magnetic energy with respect to the stellar wind push, the
harder is to open up the closed field structure, resulting in a smaller
open flux fraction. While in ISW cases
the amount of open flux is determined by the stellar wind push only,
in SDI systems it also depends on the position of the accretion
spots, since the star-disk differential rotation has the tendency to
open up all the magnetic surfaces that are not mass-loaded by the
columns\footnote{Some of these field lines open up intermittently,
producing, in our framework, the MEs phenomenon. These magnetic
surfaces are not included in our $\Phi_{open}$ estimate.}.
While this effect is likely responsible for the larger open flux
measured in SDI simulations, the interpretation of the decrease of the
open flux with the $\Upsilon$ parameter in SDI systems is less
straightforward. In our SDI cases, the increase of the $\Upsilon$
parameter corresponds also to an increase of the position of $R_t$
and, correspondingly, of the $\Upsilon_{acc}$ parameter (see Table
\ref{tab_data_sim}). A larger truncation
radius tends to displace the accretion spots to higher latitudes and
therefore to reduce the amount of fractional open flux. With our
limited set of simulations it is not possible to determine which
effect, the stellar wind push (i.e. the $\Upsilon$ value) or the
position of the truncation radius and the accretion spots (i.e. the
$\Upsilon_{acc}$ value), is more important to determine the scaling of
the open magnetic flux. In particular, since in our simulations we
changed the dipolar field intensity $B_\ast$ only, the $\Upsilon$ and
$\Upsilon_{acc}$ parameters increase (decrease) at the same time, both
contributing to the decrease (increase) of the fractional open
flux. Clearly a wider parameter space exploration, with $\Upsilon$ and
$\Upsilon_{acc}$ varying independently, is required to address this
issue.

On the other hand, we already mentioned that the presence of MEs
confines the stellar wind inside an hourglass-shaped magnetic flux
tube, instead of letting the wind expand freely as in the ISW cases.  
To better quantify this effect we plot in the right panel of Fig. \ref{fig_vprof_area}
the area of the stellar wind flux tube $S_{sw}$ versus the radial
distance from the star for two SDI and ISW cases with the same
$B_\ast/B_0$ value. Obviously, the radial dependence of $S_{sw}$ reflects
the magnetic field topology since, due to magnetic flux conservation,
$S_{sw} \propto B_{sw}^{-1}$. In both cases the magnetic field close
to the star ($R \lesssim 4R_{\ast}$) tends to keep its dipolar
potential topology, providing $S_{sw} \propto R^n$, with $n=3$. Note,
however, that at the stellar surface the area of the stellar wind is
slightly larger in the SDI case than in the ISW case, confirming that
the SDI case is characterized by a larger open magnetic flux. It
should be mentioned that the areal difference at the stellar surface
between the two stellar-wind regions corresponds to a
$\theta$-coordinate difference of a few degrees on each 
hemisphere. Nevertheless, as shown in Fig. \ref{fig_opflux_y}, such
a small difference is capable to increase the open flux by $\sim 20\%$
For low values of the wind magnetization (with $\Upsilon < 10^5$), these
angles can differ by up to $10^{\circ}$ on each hemisphere, which
results in almost a factor of two increase of
$\Phi_{open}/\Phi_{\ast}$ for SDI cases.
On a large scale ($R > 10 R_\ast$), the ISW completely opens the
magnetosphere, fills the entire domain and propagates almost radially,
with $n=2$. The stellar wind of the SDI cases remains confined within
a smaller area that expands sub-radially, with $n < 2$. 

The flux tube actually acts as a nozzle and strongly modifies the
acceleration profile of the flow. Since we are dealing with mainly
thermally driven winds (rotation and Lorentz forces are almost
negligible), the faster expanding nozzle, the ISW one, should
determine a faster decline of the thermal pressure and a faster
acceleration of the wind. This is confirmed by the left panel of 
Fig. \ref{fig_vprof_area}, where the radial profiles of the wind
velocity, the slow-magnetosonic and Alfv\'{e}n speeds of the two cases
are plotted. These quantities are averaged in time and space (along
the $\theta$ coordinate). Within a few stellar radii the
three speed profiles are quite similar in both cases and the two
outflows reach the slow-magnetosonic point around $R \approx 4 R_\ast$. 
Confirming our hypothesis, on a larger scale the ISW case accelerates
more rapidly and becomes noticeably faster than the SDI
case. Moreover, since the magnetic field is compressed within a
smaller area, the SDI Alfv\'{e}n speed becomes larger than the ISW
one. As a consequence, while the ISW becomes super-Alfv\'{e}nic at a
distance $R \approx 40 R_\ast$, the stellar wind of the SDI case has
not reached the Alfv\'{e}n point within our computational domain. Even
in SDI tests performed with a radial domain twice as large, the
stellar wind did not reach the Alfv\'{e}n surface. Notice that a large
difference of the radial distance from the star of the Alfv\'{e}n surface
does not automatically reflects a comparable difference of average
lever arm $\langle r_{A}\rangle$. In Appendix \ref{app_sw_torq} we
propose a simple relation between the cylindrical Alfv\'{e}n radius
$\langle r_{A}\rangle$ and the average radial distance of the
Alfv\'{e}n surface $\bar{R}_A$, 
\begin{equation}
\langle r_{A}\rangle = \bar{R}_A \sin\left(\frac{\theta_{oA}}{2}\right) \, ,
\end{equation}
where $\theta_{oA}$ is the opening angle of the Alfv\'{e}n
surface. Since for an ISW $\theta_{oA} \approx \pi/2$, the lever arm
difference is strongly reduced by the smaller opening angle of the SDI
stellar winds. 

Following the approach adopted in \citet[][see also
\citet{Kawaler:1988aa,Tout:1992aa,Matt:2008aa,Reville:2015ab}]{Pantolmos:2017aa}, 
we tried to quantify the impact of these effects (the different amount
of open flux and speed profile) and derived a simple analytic
expression for the ratio of the lever arms $\langle r_{A}\rangle$ in
two SDI and ISW cases (see Appendix \ref{app_sw_torq} for a full
derivation): 
\begin{equation}
\frac{\langle r_A^{sdi} \rangle}{\langle r_A^{isw} \rangle} =
\left[
\frac{\upsilon_{sw,A}^{isw}}{\upsilon_{sw,A}^{sdi}}
\frac{\left(\Phi_{open}/\Phi_{\ast}\right)_{sdi}^2}{\left(\Phi_{open}/\Phi_{\ast}\right)_{isw}^{2}}
\frac{\Upsilon^{sdi}}{\Upsilon^{isw}}
\right]^{1/2} = 
\left[
\frac{\upsilon_{sw,A}^{isw}}{\upsilon_{sw,A}^{sdi}}
\frac{\Upsilon^{sdi}_{open}}{\Upsilon^{isw}_{open}}
\right]^{1/2} \, ,
\label{eq_ra_va_y}
\end{equation}
where $\upsilon_{sw,A}$ is the average speed of the flow at the
Alfv\'{e}n surface and quantifies the acceleration efficiency of the
stellar wind. According to Eq. (\ref{eq_ra_va_y}), the larger lever arm displayed by
an SDI stellar wind with the same magnetization $\Upsilon$ of an ISW
(see Fig. \ref{fig_ra_y}), can be ascribed to the larger open flux and
to the slower wind speed of the SDI case. If we compare two SDI and
ISW cases with the same $\Upsilon_{open}$ (see
Fig. \ref{fig_ra_yopen}), the larger SDI lever arm should be
determined by the slower speed (i.e. $\upsilon_{sw,A}$) only. Note
that this last point is hard to prove, since for the majority of our
simulations, we cannot extract $\upsilon_{sw,A}$ because  
the stellar outflows stay on average sub-Alfv\'{e}nic within our
computational box. Therefore we cannot verify that the offset in
Fig. \ref{fig_ra_yopen} is entirely due to differences in $\upsilon_{sw,A}$. 
More subtle geometrical effects could produce an
additional scatter of the data points. 
In addition, the different $m_o$ exponent found in ISW and SDI
systems could reflect the different wind acceleration of the two
sets of simulations. As discussed in Appendix \ref{app_sw_torq}, the
flatter speed profile observed in SDI simulations (a smaller $q$
value in Eq. (\ref{eq_va_sa})) should correspond to a larger $m_o$
exponent compared to ISWs, qualitatively consistent with our
findings.

\begin{figure}
  \centering
  \resizebox{\hsize}{!}{\includegraphics{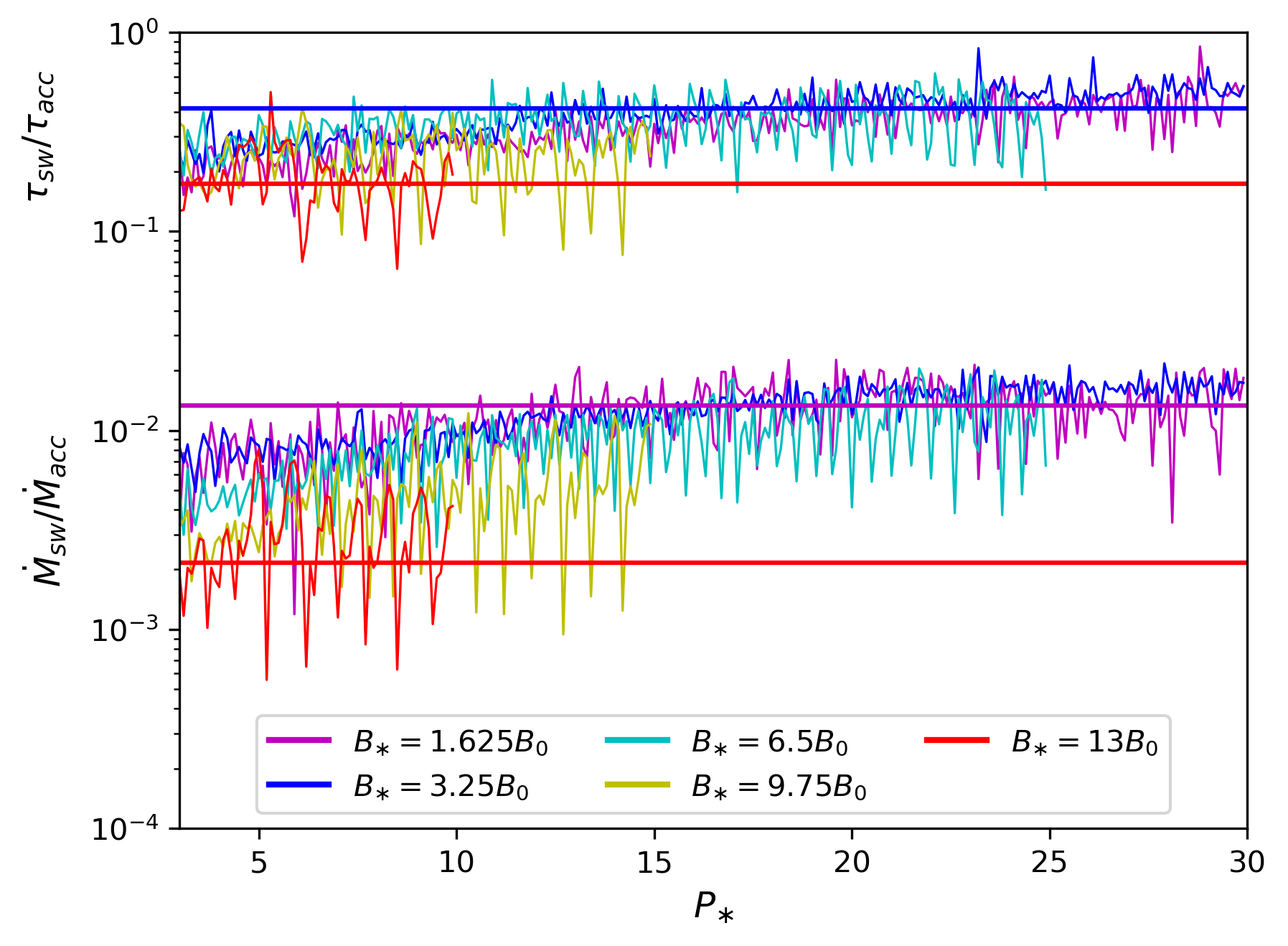}}
  \caption{Absolute values of the mass ejection efficiency
    $\dot{M}_{sw}/\dot{M}_{acc}$ and torque efficiency
    $\tau_{sw}/\tau_{acc}$ as a function of time expressed in units of
    the stellar rotation period. The straight lines in the plot
    represent the time-averaged values for the two cases exhibiting
    the lowest and highest values of these two quantities.}
  \label{fig_mdot_jdot_sw_acc}
\end{figure}

We evaluate the mass ejection and torque efficiencies of
the stellar winds by plotting in Fig. \ref{fig_mdot_jdot_sw_acc} the
time evolution of $\dot{M}_{sw}/\dot{M}_{acc}$ and
$\tau_{sw}/\tau_{acc}$. The 10th column of Table
\ref{tab_data_sim} lists the time-averaged values of
$\tau_{sw}/\tau_{acc}$. On average, the stellar outflow is able to extract
between 0.2\% (i.e., SDI case 5 with $B_{\ast} = 13\, B_0$) and 1\%
(i.e., SDI case 1 with $B_{\ast} = 1.625\, B_0$) of the mass accretion
rate. In addition, the stellar wind braking torque ranges from about
20\% (i.e., SDI case 5 with $B_{\ast} = 13 \, B_0$) to 40\%  (i.e., SDI
case 2 with $B_{\ast} = 3.25 \, B_0$) of the accretion torque. These
values are consistent with the results of
\citet{Zanni:2009aa,Zanni:2013aa}.

We close this section by presenting the functional form of the torque
scaling for stellar winds. By combining Eqs.
(\ref{eq_ra_upsilon}) and  (\ref{eq_torque_sw}) or Eqs.
(\ref{eq_ra_upsilon_open}) and (\ref{eq_torque_sw}),
we obtain
\begin{align}
\tau_{sw} &= K_{sw,s}^2 (4 \pi)^{-2m_s} \Omega_{\ast}
\upsilon_{esc}^{-2 m_s}  R_{\ast}^{2-4m_s} \dot{M}_{sw}^{1 - 2m_s}
\Phi_{\ast}^{4m_s} \label{eq_torque_sw_fit_tot}\\
&= K_{sw,o}^2 (4 \pi)^{-2m_o} \Omega_{\ast} \upsilon_{esc}^{-2 m_o}
R_{\ast}^{2-4m_o} \dot{M}_{sw}^{1 - 2m_o} \Phi_{open}^{4m_o} 
\label{eq_torque_sw_fit_open} \; ,
\end{align}
where Eq. (\ref{eq_torque_sw_fit_tot}) employs the total unsigned
surface magnetic flux $\Phi_{\ast}$, while
Eq. (\ref{eq_torque_sw_fit_open}) uses the open magnetic flux
$\Phi_{open}$. Using the corresponding values of $K_{sw,s}$,
$K_{sw,o}$ and $m_{s}$, $m_{o}$ given in Table
\ref{tab_fit_sw}, equation Eqs. (\ref{eq_torque_sw_fit_tot}) and
(\ref{eq_torque_sw_fit_open}) provide an estimate of  the magnetic
torque due to stellar winds over a wide range of surface magnetic
field strengths, both for accreting and non-accreting stars. 

\subsubsection{Magnetospheric-ejections torque}
\label{sec_me_torq}

\begin{figure}
  \centering
  \resizebox{\hsize}{!}{\includegraphics{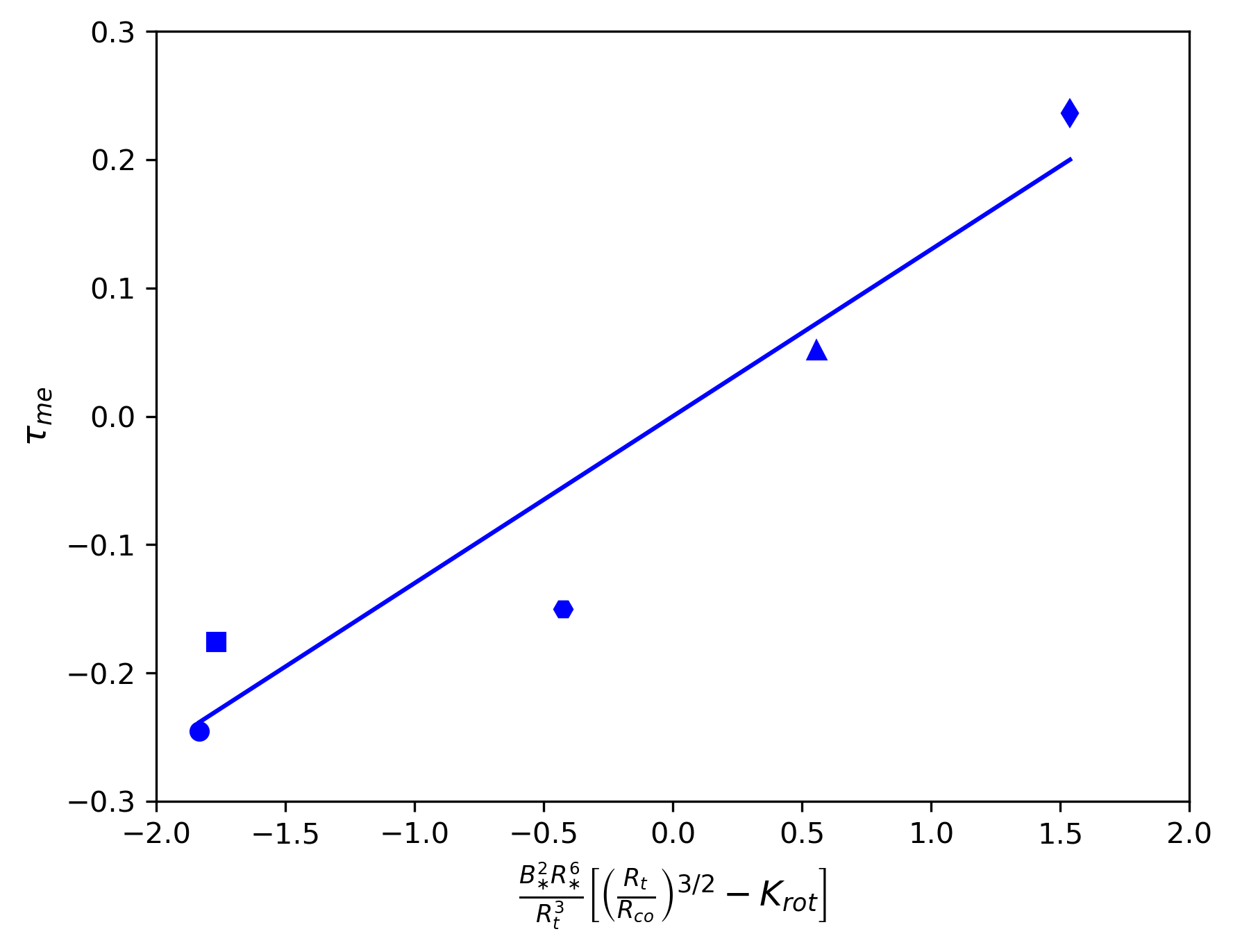}}
  \caption{Normalized $\tau_{me}$ versus quantity in curly brackets in
    Eq. (\ref{eq_torque_me_fit_func}). Symbols are the same
    as in Fig. \ref{fig_ra_y}. In the plot, a negative
    (positive) $\tau_{me}$ spins up (down) the stellar rotation. Using
    Eq. (\ref{eq_torque_me_fit_func}), the best fit to the points
    gives $K_{me} = 0.13$ and $K_{rot} = 0.46$.}
  \label{fig_jme_fit}
\end{figure}

\begin{figure}
  \centering
  \resizebox{\hsize}{!}{\includegraphics{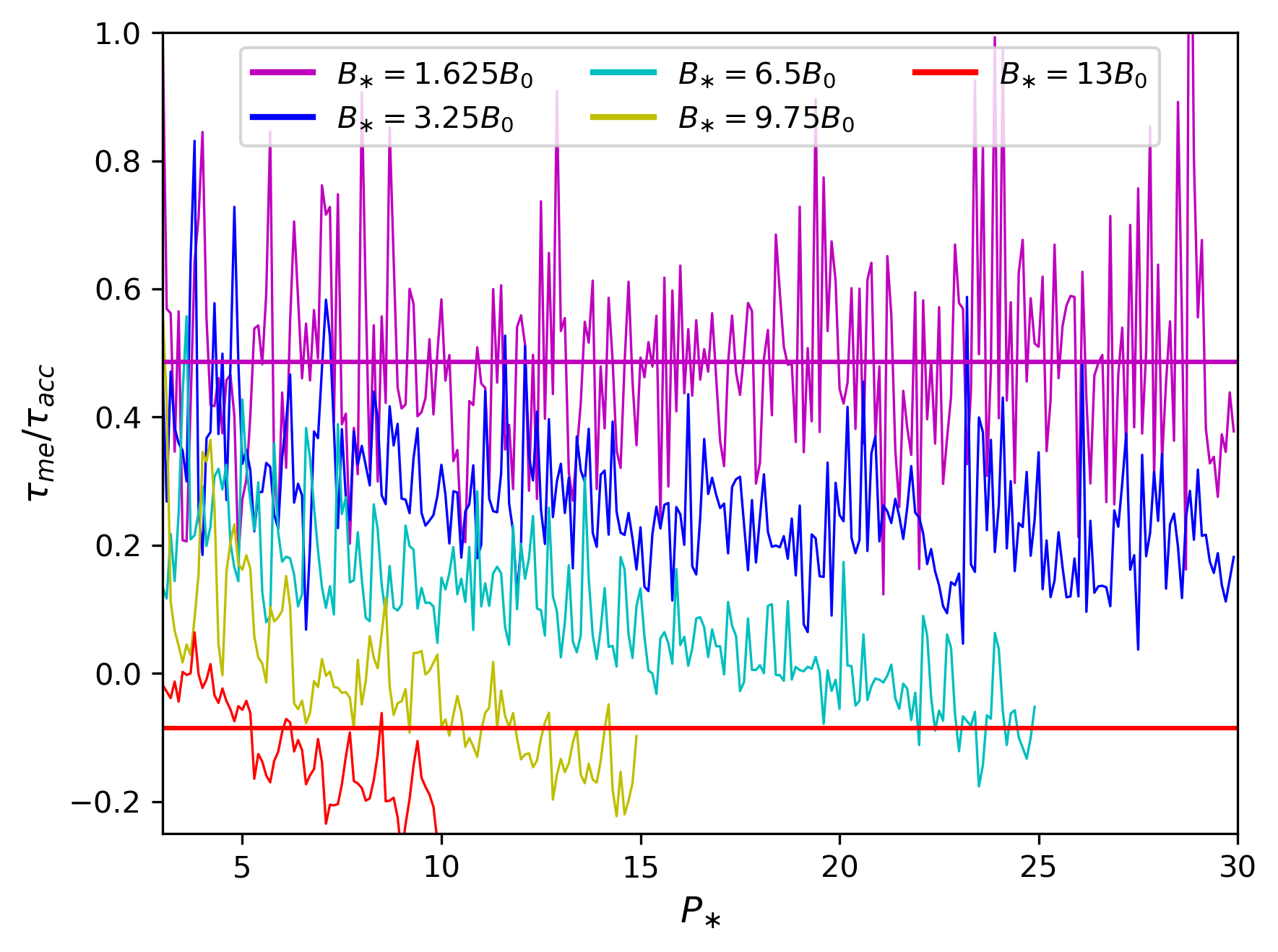}}
  \caption{$\tau_{me}/\tau_{acc}$ versus time (given in
    $P_{\ast}$). Colors and linestyles have the same meaning as in
    Fig. \ref{fig_mdot_jdot_sw_acc}. In the plot, a negative
    (positive) ratio indicates a spin-down (spin-up) torque due to
    magnetospheric ejections.}
  \label{fig_jdot_me_acc}
\end{figure}

As pointed out in \citet{Zanni:2013aa}, MEs can also directly exchange
angular momentum with the star. In that case, the direction of the
angular-momentum flow depends on the differential rotation between the 
star and the MEs. If the plasma located at the cusp of the inflated field
lines rotates faster than the stellar rotation, the star is
subject to a spin-up torque. Similarly, the star experiences a
spin-down torque if the matter rotates slower than the star.

\begin{table}[tbp]
	\caption{Best-fit coefficients of scalings for SDI simulations.}
	\label{tab_fit_sdi}
	\centering
	\resizebox{\columnwidth}{!}{
		\begin{tabular}{c c c c}
			\hline\hline
			Formulation & Coefficient & Best fit & Eq. \\
			\hline
			\multirow{3}*{$R_t/R_{\ast}$, $\tau_{acc}$} & $K_{t}$ & $1.1452
			\pm 0.0008$ & \multirow{3}*{(\ref{eq_rt_fit_func}),  (\ref{eq_torque_acc_fit_func}), (\ref{eq_torque_acc})}\\ 
			& $m_{t}$ & $0.35 \pm 0.01$ & \\
			& $K_{acc}$ & $0.79 \pm  0.01$ & \\
			\hline
			\multirow{4}*{$\langle r_{A}\rangle/R_{\ast}$, $\tau_{sw}$} & $K_{sw,s}$ & $3.8156 \pm 0.0001$ &
			\multirow{2}*{(\ref{eq_ra_upsilon}), (\ref{eq_torque_sw_fit_tot})} \\  
			& $m_{s}$ &  $0.169 \pm 0.001$  &\\
			& $K_{sw,o}$ & $ 0.495\pm0.001$  & \multirow{2}*{(\ref{eq_ra_upsilon_open}), (\ref{eq_torque_sw_fit_open})}  \\
			& $m_{o}$ & $0.439\pm0.005$ & \\ 
			\hline
			\multirow{2}*{$\tau_{me}$} & $K_{me}$ & $0.13 \pm 0.02$ & \multirow{2}*{(\ref{eq_torque_me_fit_func})}  \\
			& $K_{rot}$ & $0.46 \pm 0.02$ & \\ 
			\hline
		\end{tabular}
	}
	%
\end{table}

In order to parametrize the torque exerted by the MEs directly onto the star, 
we adopt the prescription introduced in \citet{Gallet:2019aa},
\begin{equation}
  \tau_{me} = K_{me} \left \{ \frac{B_{\ast}^2 R_{\ast}^6}{R_t^3}
  \left[ \left(\frac{R_t}{R_{co}} \right)^{3/2} - K_{rot} \right] \right \} \, ,
  \label{eq_torque_me_fit_func}
\end{equation}
where $K_{me}$ and $K_{rot}$ are dimensionless fitting coefficients.
Equation (\ref{eq_torque_me_fit_func}) assumes that the launching
region of the MEs is located close to $R_t$, so that the torque
depends on the local magnetic field strength and the differential
rotation between the truncation region and the star \citep[for a more
detailed discussion see Sect. 2.3.2 in][]{Gallet:2019aa}. In
particular, the $K_{rot}$ factor takes into account the difference
between the MEs rotation and the disk sub-Keplerian rotation around $R_t$.

The dependence of $\tau_{me}$ on the term inside the curly brackets in
the right-hand side of Eq. (\ref{eq_torque_me_fit_func}) is presented
in Fig. \ref{fig_jme_fit}. The best fit (solid line) in the plot
  has $K_{me} =  0.13$ and $K_{rot} = 0.46$ (see also Table
  \ref{tab_fit_sdi}). $K_{rot}$ is found to be less than unity,
indicative of the sub-Keplerian rotation rate of the disk 
around $R_t$, where the plasma is mass loaded along MEs field lines. In
addition, $K_{rot}$ and $K_{me}$ differ by 20\% and almost a factor of
2, respectively, compared to the values used in \citet{Gallet:2019aa}.
These differences could be a result of numerical diffusion
effects that appear in our SDI simulations. More specifically, at the
boundary between the stellar wind and MEs region there are a few
steadily open field lines anchored into the stellar surface along
which the plasma accretes close to the star and is ejected at a larger
distance, nevertheless exerting a spin-down torque onto the star. This
effect is likely due to numerical diffusion of the accretion flow into
the MEs and stellar wind regions. Despite being steadily open and
providing a spin-down torque, we choose to classify these magnetic
surfaces not as part of the stellar wind but as belonging to the MEs
region, since the conditions at the stellar surface could in principle
allow the MEs to accrete onto the stellar surface, while we strictly
require that the stellar wind extracts mass from the star.
This choice most likely tends to increase the spin-down efficiency of
the MEs and, correspondingly, to slightly underestimate the stellar
wind torques. As already discussed in Sect. \ref{sec_sdi}, this is a further
indication of the influence of the numerical subtleties on the
quantitative properties of our solutions, and the MEs in
particular.

Consistently with Eq. (\ref{eq_torque_me_fit_func}) the plot shows a
change of sign of $\tau_{me}$, going from a negative (i.e., spin up)
to a positive (i.e., spin down) value. This transition occurs at
$R_t\approx 0.6R_{co}$. The latter result agrees qualitatively with
the findings of \citet{Zanni:2013aa}.

The time-averaged values $\tau_{me}$ as a fraction of $\tau_{acc}$ are
tabulated in the 11th column of Table \ref{tab_data_sim} and the
temporal evolution of the MEs torque efficiency,
$\tau_{me}/\tau_{acc}$, is shown Fig. \ref{fig_jdot_me_acc}.
As seen in Table \ref{tab_data_sim}, the efficiency of the MEs
torque can vary between a 50\% spin-up efficiency for the lowest field
case and a 10\% spin-down efficiency for the highest field case,
changing sign in between the two. For the lower field cases ($B_{\ast}
< 5B_{0}$ or nominally $B_{\ast} < 500$ G) MEs provide a significant
contribution (between 20\% and 50\% of $\tau_{acc}$) to the spin-up
torque exerted onto the star. This regime 
requires relatively weak dipolar magnetic fields combined with a ratio
$R_t/R_{co} \lesssim 0.4$. Notice that this configuration is
compatible with many observations of classical T Tauri stars
\citep[see e.g.,][]{Johnstone:2014ab}. The average spin-down torque
for the high field cases is almost negligible, with a maximum
spin-down efficiency $\approx 10 \%$ in the highest field case
($B_{\ast} = 13B_0$, nominally corresponding to $B_{\ast} = 1.2$
kG). On the other hand Fig. \ref{fig_jdot_me_acc} shows that 
the spin-down efficiency of the MEs in the higher field cases is
increasing in time, corresponding to a decrease of the accretion rate
and the truncation radius moving closer to corotation. This result
points to the fact that in order to maximize the spin-down efficiency
of the MEs the disk should be truncated close to the corotation
radius, possibly entering a propeller regime, as suggested in
\citet{Zanni:2013aa}.

\section{Discussion}
\label{sec_disc}

In this section, we discuss the astrophysical implications of our
simulations and compare our findings with previous works from
the literature. In addition, we refer to possible limitations of our
models.

The angular momentum equation of a star rotating as a solid body is
given by
\begin{equation}
  \frac{\dot{\Omega}_{\ast}}{\Omega_{\ast}} =
  - \frac{\tau_{tot}}{J_{\ast}} - \frac{\dot{M}_{acc}}{M_{\ast}} -
  \frac{2\dot{R}_{\ast}}{R_{\ast}}.
  \label{eq_spinev}
\end{equation}
In the right hand side of Eq. (\ref{eq_spinev}), the first term
corresponds to the inverse of the characteristic spin-down/spin-up
timescale $t_{sdi} = -{J_{\ast}}/{\tau_{tot}}$ of the total exernal
torque $\tau_{tot} = \tau_{acc} + \tau_{sw} + \tau_{me}$, where
$J_{\ast} = k^2 M_{\ast}  R_{\ast}^2 \Omega_{\ast}$ is the stellar
angular momentum, with $k^2 = 0.2$ (i.e., mean radius of gyration of a fully convective star).   
The second term, $\dot{M}_{acc}/M_{\ast} \sim t_{acc}^{-1}$, is linked
to the mass-accretion timescale. This term considers the changes of
the stellar moment of inertia due to mass accretion and gives $t_{acc}
\gtrsim 10$Myr, for a typical $\dot{M}_{acc} \lesssim 10^{-7} \mathrm{M}_{\sun}\
\mathrm{yr}^{-1}$, so that it can be usually neglected. 
Finally, the third term, $-\dot{R}_{\ast}/R_{\ast} \sim
t_{KH}^{-1}$, refers to the change of the stellar moment of inertia due
to the gravitational stellar contraction and is therefore associated
with the Kelvin-Helmholtz timescale \citep[see e.g.][]{Bodenheimer:2011aa}.
From \citet{Collier-Cameron:1993aa,Matt:2010aa}, $t_{KH}$ can be
written as
\begin{equation}
  t_{KH}= 3.6 \left(\frac{M_{\ast}}{0.7M_{\sun}}\right)^2
  \left(\frac{R_{\ast}}{2R_{\sun}}  \right)^{-3}
  \left(\frac{T_{eff}}{4000 \mathrm{K}}  \right)^{-4}\ \ \mathrm{Myr} \, ,
  \label{eq_tKH}
\end{equation}
where $T_{eff}$ is the photospheric temperature.
Equation (\ref{eq_spinev}) shows that, in order to achieve a steady
stellar rotation ($\dot{\Omega}_\ast = 0$), the total external torque
should provide a net spin-down, a condition that is not verified in
any of our simulations, see Table \ref{tab_data_sim}, with a
characteristic timescale comparable to the Kelvin-Helmholtz one.  It
should be noted that stellar contraction 
on its own is able to decrease the rotation period of slow rotators
with $P_{\ast} \approx 8$ days at 1Myr down to 2 days during the disk
lifetime, taking a median value of 3 Myr, even if later studies
propose a disk lifetime for slow rotators of 9 Myr \citep[see
e.g.][]{Williams:2011aa,Gallet:2019aa}. 
Besides, in all the SDI simulations discussed in this work, the star
is subject to an external SDI torque that spins up the star providing
a characteristic spin-up timescale varying between 0.4 and 1.5Myr
(i.e. for SDI cases 5 and 1, respectively). Obviously, the
SDI stellar torques in our simulation can only provide an additional
spin-up torque to the stellar rotational evolution, they further
shorten the spin-up timescale due to contraction and clearly cannot 
explain the presence of slow rotators with $P_{\ast} \geq 8$ days at
about 10 Myr.

A possible solution to this problem could be provided by more
massive stellar winds, with a higher spin-down torque. We recall
that in our SDI models the spin-down torque of a stellar wind
ejecting less than $2\%$ of the mass accretion rate corresponds to
$20-40\%$ of the spin-up accretion torque. Despite not being
sufficient to provide an efficient enough spin-down torque, the
important result presented in this paper is that, in the range of
the parameter space explored, the stellar wind torque in accreting
systems is more efficient than in isolated stars. We found that the
presence of the accretion disk strongly influences the amount of
open flux and the shape of the wind flux tube, two main factors that
determine the wind magnetic lever-arm. We found that, for the same
dipolar field intensity $B_\ast$, the stellar wind torque in SDI
systems is 1.3 to 2.3 times higher than in ISW systems. For fixed
stellar parameters, using the SDI fits from Table \ref{tab_fit_sdi},
Eqs. (\ref{eq_torque_sw_fit_tot}) and 
(\ref{eq_torque_sw_fit_open}) show that the stellar wind torque can
be increased by a larger wind mass-loss rate. We can actually use
these scalings to estimate the wind mass ejection efficiency
required to balance at least the accretion spin-up torque in the SDI
cases presented in this paper. 
Using Eq. (\ref{eq_torque_sw_fit_tot}) we find that an ejection
efficiency $\dot{M}_{sw}/\dot{M}_{acc} \lesssim 0.05$ is needed while
Eq. (\ref{eq_torque_sw_fit_open}) provides an estimate
$\dot{M}_{sw}/\dot{M}_{acc} > 1$. The large discrepancy between these
two estimates clearly depends on the  
the different dependence of $\tau_{sw}$ on $\dot{M}_{sw}$, with the
torque increasing more rapidly with $\dot{M}_{sw}$ in
Eq. (\ref{eq_torque_sw_fit_tot}) than in
Eq. (\ref{eq_torque_sw_fit_open}). This difference is due to the fact
that using Eq. (\ref{eq_torque_sw_fit_open}), i.e. the scaling with
the wind open magnetic flux, we suppose that when the wind-mass loss
rate increases the open flux does not change, which could correspond
to a situation in which the fractional open flux is solely determined
by the position of the accretion spot and the truncation radius
(i.e. the $\Upsilon_{acc}$ value, see Eq. (\ref{eq_rt_fit_func})). In
Eq. (\ref{eq_torque_sw_fit_tot}), i.e. the scaling with the total
unsigned flux, we implicitly assume that the raise of the wind
mass-loss rate increases the fractional open flux, further amplifying
the wind lever arm and the torque (see Eqs. (\ref{eq_opflux_y}) and
(\ref{eq_ra_upsilon_upsilon_open}) in Appendix \ref{app_sw_torq}). 
This result clearly points to the fact that, in order to have a robust
estimate of the wind mass-loss rate needed to provide an efficient
enough spin-down torque, it is necessary to explore the parameter
space more extensively to have a more quantitative estimate of the
dependency of the fractional open flux on the parameters of the
system.

Anyway, requiring a high wind mass-loss rate could present some issues. We
recall that, in this work, we considered thermally-driven 
winds, that require the presence of a hot, $10^6$ K
corona. Observations show that T Tauri stars are X-ray active 
\citep[e.g.,][and references therein]{Gudel:2007aa}, which implies
the presence of million-Kelvin coronae and coronal winds during
this pre-main-sequence phase of stellar evolution
\citep[e.g.,][]{Schwadron:2008aa}. However, \citet{Matt:2007aa} showed
that thermally-driven winds should have $\dot{M}_{sw}  \lesssim
10^{-11} M_{\sun}\ \mathrm{yr^{-1}}$ for their energetics to be
compatible with the X-ray activity of CTTs, which does not seem to be
enough to provide an efficient spin-down torque. Therefore, it has
been proposed that stellar winds could be driven by other MHD
processes \citep[e.g., dissipation of Alfv\'{e}n
waves,][]{Decampli:1981aa,Scheurwater:1988aa,Cranmer:2008ab,Cranmer:2009aa}, 
possibly amplified by the impact of the accretion streams onto the
stellar surface \citep[Accretion Powered Stellar
Winds,][]{Matt:2005aa}, so as to produce more massive (with
$\dot{M}_{sw}/\dot{M}_{acc} \sim 0.1$) and colder  ($10^4$ K)
winds. The latter idea was supported by observations
\citep{Edwards:2006aa}, which suggested the presence of cool and
massive stellar winds originating from T Tauri stars, and also
confirmed by radiative-transfer calculations
\citep{Kurosawa:2006aa}. \citet{Cranmer:2008ab} tested the hypothesis
of cold stellar winds in CTTs driven by Alfv\'{e}n 
waves but found an ejection efficiency below 2\% of the mass accretion
rate. Despite the fact that the driving and energetics of CTTs stellar
winds still remain an open question, our current findings suggest that
the torque provided by magnetized stellar outflows can not
single-handedly solve the angular momentum evolution problem of CTTs. 

An additional spin-down torque could be provided by MEs. However, in
this work, MEs provided either a spin-up torque or a weak spin-down
torque. \citet{Zanni:2013aa} showed that the spin-down torque by MEs
becomes more efficient when the truncation radius approaches the
corotation radius. This is confirmed by our results, since we find an
increase of the spin-down efficiency as the truncation radius
increases. However truncating the disk close to corotation causes the
transition to a (weak) propeller regime, characterized by a strong (an
order of magnitude or more) and fast (of the order of one stellar
period) variability of the mass accretion rate, which is not observed
\citep{Costigan:2014aa,Venuti:2014aa}. The analysis presented in this
paper focused on CTTs being in a steady accretion regime and therefore, we 
leave this investigation for future studies. It should be noted that
the scenario of a propeller regime as a solution to explain the rotational
evolution of CTTs is supported by both MHD simulations
\citep{Romanova:2005aa,Romanova:2009ab,Ustyugova:2006aa,Zanni:2013aa}
and angular-momentum-evolution studies \citep{Gallet:2019aa}.

It is also important to point out that, since in our models we varied
the magnetic field strength $B_\ast$ only, the simulations displayed a
$\dot{M}_{acc} - B_\ast$ correlation, with $\dot{M}_{acc}$ increasing
with $B_\ast$, see discussion in \S\ref{sec_acc_torq}, that has no
observable counterpart. Obviously, in our simulations the accretion
rate can be changed independently of the $B_\ast$ value, by varying
for example the disk surface density and/or the $\alpha$
parameter(s).

Finally, our simulations do not consider complex field
topologies. Observations of T Tauri stars show their magnetic fields
to be multipolar (e.g., dipolar accompanied by strong octupolar
components). Complex field geometries can influence the stellar-wind
braking torque in two different ways. First, by affecting the location
of the accretion hotspot \citep{Mohanty:2008aa,Alencar:2012aa}, which can modify the
area that ejects the stellar outflow. This effect could have an impact
on the stellar-wind acceleration/speed and mass-loss rate, therefore
changing the stellar wind torque. Second, by affecting the amount of
open flux available to the wind, which has an impact on the
length of the magnetic lever arm and consequently, on the braking
torque. In particular, stellar winds with single high-order field
topologies (e.g., quadrupoles, octupoles) exhibit decreased torque
efficiency (for fixed wind energetics) since the magnetic field decays
faster with distance (or the wind carries less open flux), which
results in smaller magnetic lever arms
\citep[e.g.,][]{Reville:2015ab,Garraffo:2016aa}. 
For mixed topolgies (e.g., superpositions of dipoles and
quadrupoles/octupoles), simulations of isolated stellar winds
demonstrate that the stellar-wind torque is mainly determined by the
lowest order field topology
\citep{Finley:2017aa,Finley:2018aa}. However, supplementary analysis
by \citet{See:2019ab} showed that higher order fields (in mixed
geometries) can have an effect on stellar-wind torques if the
Alfv\'{e}n surface is reached at a distance from the stellar surface
where the higher-order-field strengths have not declined enough to be
considered negligible. Therefore, more realistic/complex field
topologies shall be considered in future SDI simulations to improve
the accuracy of the current stellar-wind torque presciptions.

\section{Conclusions}
\label{sec_concl}

In this work, we presented 2.5D, MHD, time-dependent,
axisymmetric numerical simulations of magnetized rotating stars
interacting with their environment. We focused on two 
different types of numerical solutions and split our simulations in
two sets. The first set included 5 numerical solutions of stellar winds
coming from isolated stars (ISW), representing weak-line T Tauri and
main-sequence stars. The second set included 5 star-disk-interaction
simulations (SDI), representing classical T Tauri systems. In both sets we
assumed dipolar field geometries, slow stellar rotation with a stellar
spin rate corresponding to 5\% of the break-up speed and
thermally-driven stellar winds. In the SDI simulations the disk was
taken to be viscous and resistive, using a standard alpha
prescription, with fixed values for the $\alpha$
parameters and initial disk surface density. In each set, the five
simulations are characterized by a different field strength.  
We provided parametrizations for all the external torques (due
to stellar winds, magnetospheric ejections, and accretion funnel
flows) exerted at the surface of a star mangnetically interacting with its
accretion disk. We also compared the magnetic-braking efficiency of
the stellar winds in the two distinct systems considered
here (i.e., diskless stars and star-disk interacting systems).

The following points summarize the conclusions of this work.

\begin{enumerate}

\item In star-disk interacting systems we find power-law scalings of the 
  stellar-wind Alfv\'{e}n radius with the wind total $\Upsilon$ or open
  $\Upsilon_{open}$ magnetization \citep[e.g.,][]{Matt:2012ab,Reville:2015ab,Pantolmos:2017aa,Finley:2018aa}
  akin to those found in ISW systems (see Figs. \ref{fig_ra_y},
  \ref{fig_ra_yopen}, and Eqs. \ref{eq_upsilon}, \ref{eq_upsilon_open}).
  Within the parameter space considered, our simulations showed that for a given
  value of $\Upsilon$ or $\Upsilon_{open}$ stellar winds in SDI systems
  exhibit a larger magnetic lever arm and an overall more efficient
  spin-down torque compared to stellar winds from isolated stars. We found that
  the presence of the accretion funnels tends to increase the amount
  of open magnetic flux carried by the wind, while the magnetospheric
  ejections confine the stellar wind in a narrower nozzle that
  modifies the flow acceleration/velocity profile producing slower
  stellar-wind solutions. We verified that both these factors contribute to
  increase the lever arm. In our simulation sample the stellar wind
  torque is from 1.3 to 2.3 times stronger in SDI systems than it is
  in ISW systems with the same $B_\ast$. For the relatively
  low wind mass-loss rates produced by our SDI simulations,
  corresponding to less than 2\% of the mass accretion rate, the
  stellar wind is able to extract between 20 - 40\% of the accreting
  angular momentum.
  
 \item In the SDI simulations, for the range of field strengths and mass
   accretion rates considered in this work, the disk is truncated up to 66\% 
   of the corotation radius.
   We obtain a simple power-law scaling of the truncation radius
   $R_t$ with the accretion-related magnetization parameter
   $\Upsilon_{acc}$ (see Eq. \ref{eq_rt_yacc}) and we further verify
   that the spin-up torque due to accretion is linearly proportional
   to the mass accretion rate times the disk specific angular momentum
   at the truncation region (see Eq.\ref{eq_torque_acc}).

 \item Following \citet{Gallet:2019aa}, we modeled the magnetic torque
   exterted by MEs on the star as a differential rotation effect between
   the star and the MEs and we derived a torque parametrization over a
   range of stellar field strengths (see
   Eq. \ref{eq_torque_me_fit_func}). In our set of simulations, MEs
   provide either a spin-up torque, up to $\sim$50\% of the accretion
   one, or a weak spin-down torque, up to $\sim$10\% of the accretion
   one, depending on the relative position of the truncation radius
   $R_t$ with respect to the corotation radius $R_{co}$. The
   transition from a spin-up to a spin-down MEs torque occurs at
   $R_t\approx 0.6R_{co}$.
     
\item In all the SDI cases analyzed, the star is subject to a net spin-up
  torque. The latter result yields a spin-up timescale due to the sum of all
  external torques of about 1 Myr, which is even shorter than
  the spin-up timescale due to stellar contraction (see Sect. \ref{sec_disc}
  for more details). Therefore, the stars simulated in our
  study would have the tendency to spin-up rather rapidly and, without
  the presence of a more efficient spin-down mechanism, they could not 
  explain the existence of slow rotators (with $P_{\ast} \gtrsim 8$ days)
  after about 10 Myr. We argued that a stellar wind with a mass-loss rate
  higher than the one considered in our models could increase the
  spin-down efficiency. However, a more extensive exploration of the
  parameter space is necessary in order to provide robust
  estimates on the stellar-wind ejection efficiency needed to
  counteract the spin-up torques. Finally, our simulations showed
  that, as the truncation radius gets closer to corotation, with the
  system moving towards a propeller regime, the spin-down efficiency
  of the MEs increases, further contributing to the solution of the
  rotational evolution problem of CTTs. The latter agrees with the
  conclusions of other previous studies
  \citep{Romanova:2005aa,Romanova:2009ab,Ustyugova:2006aa,Zanni:2013aa,Gallet:2019aa}.
  
\end{enumerate}

\begin{acknowledgements}
  The authors thank Catherine Dougados, Sean Matt, Lewis Ireland, and 
  Florian Gallet for helpful discussions during the production of the
  manuscript. This project has received funding from the European
  Research Council (ERC) under the European Union’s Horizon 2020
  research and innovation programme (grant agreement No 742095; {\it
    SPIDI}: Star-Planets-Inner Disk-Interactions)”;
  https:www.spidi-eu.org. All the figures within this work were
  produced using the Python-library Matplotlib \citep{Hunter:2007aa}.
  
\end{acknowledgements}


\bibliographystyle{aa}


\begin{appendix}

  \section{The equation of state}
  \label{app_eos}

We assume the thermal equation of state of an ideal gas $PV = Nk_BT$,
where $P$ is the pressure, $V$ is the volume occupied by the gas, $N$
is the number of molecules, $k_B$ is the Boltzmann constant and $T$ is
the temperature. In terms of the mass density $\rho = N m /V$, $P=\rho
k_B T/m$ where $m$ is the mean molecular mass. To simplify the
notation, assuming a constant mean molecular mass, we include $k_B$
and $m$ in the definition of the ``temperature'' so that $P=\rho
T$. With this notation the temperature $T$ has the c.g.s. dimensions
of $\mathrm{cm}^2 \, \mathrm{s}^{-2}$, i.e. the square of a speed. We
write the first law of thermodynamics:

\[
\mathrm{d}U = T\mathrm{d}S -P\mathrm{d}V \quad ,
\]
where $U$ is the internal energy and $S$ is the entropy, as:
\begin{equation}
\label{eq:fpt}
\mathrm{d}u = T\left( \mathrm{d}s + \frac{\mathrm{d}\rho}{\rho} \right) \quad ,
\end{equation}
where $u = U/M$ and $s = S/M$, with the gas mass $M = N m = \rho V$,
are the specific, i.e. per unit mass, internal energy and entropy
respectively. We also take the caloric equation of state for an ideal
gas:

\begin{equation}
\label{eq:caleos}
\mathrm{d} u = C_V \mathrm{d} T \qquad \mathrm{d} h = C_p \mathrm{d} T \quad ,
\end{equation} 
where $h = u + T$ is the specific enthalpy and $C_V$ and $C_p$ are the
specific heat capacities at constant volume and pressure
respectively. With our definition of temperature, $C_V$ and $C_p$ are
adimensional and the Mayer's relation becomes $C_p-C_V = 1$. For a
calorically perfect ideal gas $C_p$ and $C_V$ are constant and equal
to $C_p = \gamma/\left(\gamma-1\right)$ and $C_V =
1/\left(\gamma-1\right)$ where $\gamma = C_p/C_V$ is their ratio. We model the gas as thermally but not calorically perfect, i.e. its heat
capacities can depend on temperature, and we assume a piecewise
constant/linear dependency:

\begin{equation}
C_p =
\begin{cases}
C_{p0} & \mathrm{for} \; T \leq T_0\\
\left(C_{p1}-C_{p0}\right)\frac{T-T_0}{T_1-T_0}+C_{p0} & \mathrm{for} \; T_0 < T < T_1 \quad , \\
C_{p1} & \mathrm{for} \; T \geq T_1
\end{cases}
\end{equation}
where we take $C_{p0} = \gamma_0/(\gamma_0-1)$ and $C_{p1} =
\gamma_1/(\gamma_1-1)$, with $\gamma_0$ and $\gamma_1$ specifying the
ratio of the heat capacities for $T<T_0$ and $T>T_1$ respectively. Integrating
Eq. (\ref{eq:caleos}) we can write the specific enthalpy as:
\begin{equation}
\label{eq:enth}
h = 
\begin{cases}
C_{p0} T & \hspace{-2cm} \mathrm{for} \; T \leq T_0\\
\left[ 0.5\left(C_{p1}-C_{p0}\right)X^2+C_{p0}X\right]\left(T_1-T_0 \right) + C_{p0}T_0 &                                \\
          & \hspace{-2cm}\mathrm{for} \; T_0 < T < T_1 \\
C_{p1} T + 0.5\left(C_{p0} -C_{p1}\right) \left(T_1+T_0\right) & 
\hspace{-2cm} \mathrm{for} \; T \geq T_1
\end{cases}
\end{equation}
where $X = (T-T_0)/(T_1-T_0)$. The specific internal energy is defined
as $u = h-T$. Using Eq. (\ref{eq:fpt}) and (\ref{eq:caleos}) we can
derive the sound speed $c_s$
\[
c_s^2 = \left. \frac{\partial P}{\partial \rho} \right\vert_s = \gamma \frac{P}{\rho} \quad ,
\]
where $\gamma = C_p/(C_p-1) = C_p/C_V$. From the first law of
thermodynamics Eq. (\ref{eq:fpt}) we can also derive the expression
for the specific entropy $s$:

\begin{equation}
\label{eq:entr}
s = s_0 + \log\left[ \frac{f\left( T \right)^\frac{1}{\gamma_0-1}}{\rho} \right] \quad ,
\end{equation}
where $s_0$ is an integration constant and $f(T)$ is:
\begin{equation*}
f\left(T\right) =
\begin{cases}
T & \hspace{-1cm} \mathrm{for} \; T \leq T_0\\
T \left(\frac{T}{T_0}\right)^\frac{\left(C_{p0}
    -C_{p1}\right)T_0}{\left(C_{p0} - 1\right) \left(T_1-T_0\right)}
\exp\left(\frac{C_{p1}-C_{p0}}{C_{p0}-1} X \right) & \hspace{-1cm}
\mathrm{for} \; T_0 < T < T_1 \\ 
T \left(\frac{T}{T_1} \right)^\frac{C_{p1}-C_{p0}}{C_{p0}-1} \left(
  \frac{T_1}{T_0}\right)^\frac{\left(C_{p0}
    -C_{p1}\right)T_0}{\left(C_{p0} - 1\right) \left(T_1-T_0\right)}
\exp\left(\frac{C_{p1}-C_{p0}}{C_{p0}-1} \right) &  \\ 
& \hspace{-1cm} \mathrm{for} \; T \geq T_1\\
\end{cases}
\end{equation*}
Since the specific entropy in the absence of irreversible processes
(dissipative heating, cooling) obeys a scalar equation:
\[
\frac{\partial \rho s}{\partial t} + \nabla \cdot \left( \rho s \vec{v} \right) = 0 \quad ,
\] 
it is possible to redefine the specific entropy as an arbitrary
function of Eq. (\ref{eq:entr}). Since the $\log$ and $\exp$
functions, which would be extensively used to derive entropy from 
temperature and temperature from entropy respectively, are 
computationally expensive, we use the simpler expression:

\begin{equation}
\label{eq:entrnolog}
s = \frac{f\left(T\right)}{\rho^{\gamma_0-1}} \quad .
\end{equation}
In our SDI simulations we assume $T_0 = 0.01 \; (GM_\star/R_\star)$,
$\gamma_0 = 5/3$, $T_1=0.1\; (GM_\star/R_\star)$, $\gamma_1 = 1.05$.

\section{Analytic scaling of the stellar-wind Alfv\'{e}n radii}
\label{app_sw_torq}

At the Alfv\'{e}n surface of an MHD flow the wind speed,
$\upsilon_{sw,A}$,  equals the local Alfv\'{e}n speed, $\upsilon_{A}$
\begin{equation}
  \upsilon_{sw,A}^2 = \upsilon_{A}^2 = {B_A^2 \over 4 \pi \rho_{sw,A}},
  \label{eq_vwind_va}
\end{equation}
where $\rho_{sw,A}$ and $B_{A}$ are the local plasma density, and
magnetic field strength, respectively. Approximating the stellar-wind
open magnetic flux to be $\Phi_{open} \sim B_{A}S_A$, where $S_A$ is 
the area of the Alfv\'{e}n surface, Eq. (\ref{eq_vwind_va}) can
be manipulated to yield

\begin{equation}
  S_A =\frac {\Phi_{open}^2}{4 \pi \dot{M}_{sw} \upsilon_{sw,A}}.
  \label{eq_sa_open_flux}
\end{equation}
Using Eqs. (\ref{eq_upsilon}) and (\ref{eq_upsilon_open})
to define the magnetizations $\Upsilon$ and $\Upsilon_{open}$, Eq.
(\ref{eq_sa_open_flux}) can be rewritten as
\begin{equation}
  \frac{S_A}{R_{\ast}^2} = \frac{\upsilon_{esc}}{\upsilon_{sw,A}}
    \Upsilon_{open} = \frac{\upsilon_{esc}}{\upsilon_{sw,A}}
      \left( \frac{\Phi_{open}}{\Phi_{\ast}} \right)^2 \Upsilon \, .
  \label{eq_sa_yopen_y}
\end{equation}
We approximate $S_A$ as
\begin{equation}
  S_A = 4 \pi \bar{R}_A^2(1 -  \cos\theta_{oA}) ,
  \label{eq_sa_costh}
\end{equation}
where $\bar{R}_A$ is the average spherical radius of the Alfv\'{e}n surface
and $\theta_{oA}$ is the opening
angle, measured from the rotation axis, of the outflow flux tube at the
Alfv\'{e}n surface. As discussed in this paper, for ISW cases this angle is
about $90^{\circ}$, since the flow completely opens the stellar
magnetic field (see e.g. Fig. \ref{fig_rho_isw}). For the SDI
cases, $\theta_{oA}$ is smaller, typically less than $45^{\circ}$ (see e.g.
Fig. \ref{fig_rho_sdi}), due to the funnel-shaped geometry of the
outflow. 
If we approximate the average cylindrical radius of the Alfv\'{e}n
surface $\bar{r}_A$ as 
\begin{equation}
  \bar{r}_A = \bar{R}_A \sin\left(\frac{\theta_{oA}}{2}\right) \, ,
\end{equation}
we obtain from Eq. (\ref{eq_sa_costh}),
\begin{equation}
   S_A = 4\pi \bar{R}_{A}^2 \left(1 -\cos\theta_{oA}\right) = 8 \pi
   \bar{R}_{A}^2 \sin^2\left(\frac{\theta_{oA}^{sdi}}{2}\right)
   \propto \bar{r}_A^2 ,
  \label{eq_ra_sw_app}
\end{equation}
where $\bar{r}_A$ represents the average cylindrical radius of the
Alfv\'{e}n surface, considering only its geometrical properties. If we
assume that $\bar{r}_A = \langle r_A \rangle$, i.e. the effective
Alfv\'{e}n radius, this equation suggests that $S_A \propto  \langle
r_A \rangle^2$ independently of the opening angle of the wind, either
fully open, as in an isolated star, or confined in a smaller funnel,
as in SDI systems. 
Following for example \citet{Reville:2015ab,Pantolmos:2017aa}, we
assume that $\upsilon_{sw,A}/\upsilon_{esc}$ scales as a power law
with $S_A$
\begin{equation}
\upsilon_{sw,A}^2 \propto \upsilon_{esc}^2
\left(\frac{S_A}{R_{\ast}^2} \right)^q \, ,
\label{eq_va_sa}
\end{equation}
where we suppose that the speed of the plasma at the 
Alfv\'{e}n surface is proportional to the specific energy injected in
the wind, expressed as a function of the escape speed, times the
expansion rate of the flow. The exponent $q$ is determined
by the local acceleration of the wind at $S_A$.
Combining Eqs. (\ref{eq_sa_yopen_y}), (\ref{eq_ra_sw_app}) and
(\ref{eq_va_sa}), we can recover the scaling
Eq. (\ref{eq_ra_upsilon_open}) 
\begin{equation}
\label{eq_ra_upsilon_open_app}
\frac {\langle r_{A}\rangle}{R_{\ast}} = K_{sw,o} \Upsilon_{open}^{m_o} \, ,
\end{equation}
where the exponent $m_o = 1/(2+q)$ should depend on the speed profile
of the wind at the Alfv\'{e}n surface. If we further assume a
power-law scaling for 
$\Phi_{open}/\Phi_{\ast}$ with parameter 
$\Upsilon$, 
\begin{equation}
\frac{\Phi_{open}}{\Phi_{\ast}} \propto \Upsilon^n \, ,
\label{eq_opflux_y}
\end{equation}
we can also recover the scaling Eq. (\ref{eq_ra_upsilon})
\begin{equation}
\label{eq_ra_upsilon_upsilon_open}
\frac {\langle r_{A}\rangle}{R_{\ast}} = K_{sw,o} \Upsilon_{open}^{m_o} =
K_{sw,o} \left[\left(\frac{\Phi_{open}}{\Phi_{\ast}}\right)^2\Upsilon\right]^{m_o} = K_{sw,s} \Upsilon^{m_s} \, ,
\end{equation}
where $m_s = m_o(2n+1)$. Clearly, while
Eq. (\ref{eq_ra_upsilon_open_app}) depends on the shape of the wind
flux tube and the speed profile at the Alfv\'{e}n surface,
Eq. (\ref{eq_ra_upsilon_upsilon_open}) depends also on the fractional
open flux dependence on $\Upsilon$. Notice that, while the scaling
Eq. (\ref{eq_opflux_y}) represents a reasonable assumption for
isolated stellar winds, in SDI systems the stellar open flux could
also depend on other factors, most notably the mass accretion rate and
the $\Upsilon_{acc}$ accretion parameter.  

In order to directly compare the Alfv\'{e}n radii of stellar winds in
isolated and accreting stars, without taking into account the exact
speed profile Eq. (\ref{eq_va_sa}) or the scaling of the fractional
open flux Eq. (\ref{eq_opflux_y}), we can combine
Eqs. (\ref{eq_sa_yopen_y}) and (\ref{eq_ra_sw_app}) to get
\begin{equation}
  \frac{\langle r_A^{sdi} \rangle}{\langle r_A^{isw} \rangle} =
  \left[ \frac{\upsilon_{sw,A}^{isw}}{\upsilon_{sw,A}^{sdi}}
    \frac{\left(\Phi_{open}/\Phi_{\ast}\right)_{sdi}^2}{\left(\Phi_{open}/\Phi_{\ast}\right)_{isw}^{2}}
    \frac{\Upsilon^{sdi}}{\Upsilon^{isw}} \right]^{1/2} =
  \left[ \frac{\upsilon_{sw,A}^{isw}}{\upsilon_{sw,A}^{sdi}}
    \frac{\Upsilon^{sdi}_{open}}{\Upsilon^{isw}_{open}} \right]^{1/2}
  \label{eq_ra_va_y_ap} 
\end{equation}

\end{appendix}

\end{document}